\documentclass[11pt,a4paper]{article}
\usepackage{jheppub}
\usepackage{bm}
\usepackage{amsmath}	

\def\xiG{\xi_G}
\def\xiW{\xi_W}
\def\xiB{\xi_B}

\def\lam{\ensuremath{\hat\lambda}}

\newcommand{\ala}{\ensuremath {a_1}}
\newcommand{\alb}{\ensuremath {a_2}}
\newcommand{\alc}{\ensuremath {a_s}}

\newcommand{\eps}{{\epsilon}} 

\newcommand{\ep}{\ensuremath{\epsilon}} 
\newcommand{\MS}{{\ensuremath{\overline{\mathrm{MS}}}}}

\def\T2F{{T^{\,2}_{\! F}}}

\def\Dhat{\ensuremath{\hat{D}}}

\def\qV{\ensuremath{\tilde V}}
\def\qG{\ensuremath{\tilde G}}
\def\qW{\ensuremath{\tilde W}}
\def\qB{\ensuremath{\tilde B}}
\def\bgV{\ensuremath{\hat V}}
\def\bgG{\ensuremath{\hat G}}
\def\bgW{\ensuremath{\hat W}}
\def\bgB{\ensuremath{\hat B}}

\def\LG{\ensuremath{\mathcal{L}_{\mathrm{G}}}}
\def\LH{\ensuremath{\mathcal{L}_{\mathrm{H}}}}
\def\LF{\ensuremath{\mathcal{L}_{\mathrm{F}}}}
\def\LFP{\ensuremath{\mathcal{L}_{\mathrm{FP}}}}
\def\LGF{\ensuremath{\mathcal{L}_{\mathrm{GF}}}}

\DeclareMathOperator{\tr}{tr}
\newcommand{\YtD}{\ensuremath {\mathcal{Y}_{d}}}
\newcommand{\YtU}{\ensuremath {\mathcal{Y}_{u}}}
\newcommand{\YtL}{\ensuremath {\mathcal{Y}_{l}}}

\newcommand{\YtDD}{\ensuremath {\mathcal{Y}_d^2}}
\newcommand{\YtUU}{\ensuremath {\mathcal{Y}_u^2}}
\newcommand{\YtLL}{\ensuremath {\mathcal{Y}_l^2}}

\newcommand{\YtDs}{\ensuremath {\mathcal{Y}_{dd}}}
\newcommand{\YtUs}{\ensuremath {\mathcal{Y}_{uu}}}
\newcommand{\YtLs}{\ensuremath {\mathcal{Y}_{ll}}}

\newcommand{\YtUD}{\ensuremath {\mathcal{Y}_{ud}}}

\newcommand{\NR}{\ensuremath {N_c}}

\newcommand{\NY}{\ensuremath {n_Y}}
\newcommand{\NGen}{\ensuremath {n_G}}
\newcommand{\ItoR}{\ensuremath {T_F}}
\newcommand{\cA}{\ensuremath {C_A}}
\newcommand{\cR}{\ensuremath {C_F}}
\newcommand{\zetathree}{\ensuremath {\zeta(3)}}

\renewcommand{\text}[1]{#1}
\def \eps {\epsilon}
\def \invep {\frac{1}{\epsilon}}
\def \epp {\frac{1}{\epsilon^2}}
\def \eppp {\frac{1}{\epsilon^3}}

\title{Anomalous dimensions of gauge fields and gauge coupling
	beta-functions in the Standard Model at three loops}
\author[a]{A.~V.~Bednyakov,}
\author[a]{A.~F.~Pikelner}
\author[b]{and V.~N.~Velizhanin}
\affiliation[a]{Joint Institute for Nuclear Research,\\
  141980 Dubna, Russia}
\affiliation[b]{Theoretical Physics Department, Petersburg Nuclear Physics Institute,\\ 
  Orlova Roscha, Gatchina, 188300 St.~Petersburg, Russia}
\emailAdd{bednya@theor.jinr.ru}
\abstract{We present the results for three-loop gauge field anomalous dimensions in the SM
	calculated in the background field gauge within the unbroken phase of the model.
The results are valid for the general background field gauge parameterized by three independent parameters.
Both quantum and background fields are considered.
The former are used to find three-loop anomalous dimensions for the gauge-fixing parameters, and
	the latter allow one to obtain the three-loop SM gauge beta-functions.
Independence of beta-functions of gauge-fixing parameters serves as a validity
	check of our final results.}
\keywords{Renormalization Group, Standard Model}
\arxivnumber{1210.6873}

\begin{document}
\maketitle

\section{Introduction}

In spite of the fact that the Standard Model has many unsatisfactory aspects Nature still does not
	allow us to find some solid evidence for the existence  of a more fundamental theory with new particles
	and/or interactions.
Due to the joint efforts of both experimentalists and theoreticians we are about to enter the only
	unexplored part of the SM and unveil the mechanism of electroweak symmetry breaking.
According to the recent experimental results, there is strong evidence for the existence of the Higgs boson,
the last missing ingredient of the SM spectrum \cite{:2012gk,:2012gu}.

The mass of the higgs seems to be located at the boundary of the so-called stability and instability regions \cite{Krasnikov:1978pu,Hung:1979dn,Politzer:1978ic}
	in the SM phase diagram (see Refs.~\cite{Bezrukov:2012sa,Degrassi:2012ry,Alekhin:2012py} for recent studies).
This fact implies that the SM can be potentially valid
	up to a very high scale (e.g., Plank scale).

In this situation, it is important to know how the running SM parameters evolve with energy scale.
The analysis of high energy behavior is usually divided into two parts.
The first one is the determination of running \MS-parameters from some (pseudo)observables.
This procedure is usually referred to as  ``matching''.
The second one utilizes renormalization group equations (RGEs) to find the corresponding values
at some ``New Physics'' scale.
In order to carry out such an analysis consistently one usually use $(L-1)$-loop matching to find
boundary conditions for $L$-loop RGEs (see, e.g., \cite{Collins:1984xc}).
It is worth pointing that the  advantage of the minimal-subtraction prescription lies in the fact
that one needs to know only the ultraviolet (UV) divergent part of all the required diagrams.
The latter has a simple polynomial structure in mass and momenta (once sub\-divergences are subtracted).
Due to this, \MS~beta functions and anomalous dimensions can be  relatively easily extracted
from Green functions by solving a single scale problem with the help of the so-called infrared rearrangements (IRR)
\cite{Vladimirov:1979zm}.

One-
and two-loop results for SM beta functions have been known for quite a long time
\cite{Gross:1973id,Politzer:1973fx,Jones:1974mm,Tarasov:1976ef,Caswell:1974gg,Egorian:1978zx,Jones:1981we,Fischler:1981is,Machacek:1983tz,Jack:1984vj,Gorishnii:1987ik}
and are summarized in \cite{Arason:1991ic}.
Until recently,  three-loop corrections were known only partially
\cite{Curtright:1979mg,Jones:1980fx,Tarasov:1980au,Larin:1993tp,Steinhauser:1998cm,Pickering:2001aq}.

Having a well tested method for calculation of three-loop renormalization
	constants~\cite{Velizhanin:2008rw,Velizhanin:2012nm,Bagaev:2012bw}
	and an experience in the calculations in the Standard Model and its minimal
	supersymmetric extension \cite{Bednyakov:2007vm,Bednyakov:2009wt,Bednyakov:2010ni}
	we are planning to perform the calculation of all renormalization group coefficients in the
	third order of perturbation theory extending the results of
	Refs.~\cite{Machacek:1983tz,Machacek:1983fi,Machacek:1984zw} to one
	more loop.

In this paper, we present our first step in this direction: the results for three-loop anomalous dimensions of the SM gauge fields.
Since we are only interested in UV-divergences for the fields and dimensionless parameters, 
	we do not consider the effects related to spontaneous breaking of electroweak symmetry
	and, as a consequence, can neglect all dimensionful parameters of the model.
Moreover, we made use of the background-field gauge (BFG)
	(see, e.g., Refs.~\cite{Abbott:1980hw,Abbott:1981ke}) to carry out our calculation.
In this gauge, due to the simple QED-like Ward identities 
	involving background fields,
	one can easily obtain expressions for the beta-functions by considering 	
	the two-point functions with external background particles.

	During the work on this project a few papers on the same topic appeared
	\cite{Mihaila:2012fm,Mihaila:2012pz} (gauge couplings)
	and \cite{Chetyrkin:2012rz} (Top Yukawa and higgs self-interactions).
	Since the authors of \cite{Mihaila:2012fm,Mihaila:2012pz} carried out a similar calculation, let us mention
	that our setup differs from that used in Ref.~\cite{Mihaila:2012fm,Mihaila:2012pz} in several aspects.

	Firstly, for the diagram generation we solely rely on \texttt{FeynArts} \cite{Hahn:2000kx}.
Since the diagrams are evaluated with the help of the \texttt{MINCER} package~\cite{Gorishnii:1989gt},
	a mapping to the \texttt{MINCER} notation for momenta is required.
This problem was solved by hand with the help of the \texttt{DIANA}~\cite{Tentyukov:1999is} topology files
	which were prepared during our previous calculations \cite{Velizhanin:2008rw}.
Based on these files a simple script was written which allows one to perform the mapping between the \texttt{FeynArts}
	and \texttt{MINCER} notation\footnote{During the preparation of the final version of the paper a routine was written that automatically maps the \texttt{FeynArts} topologies onto that of \texttt{MINCER}.}.

	Secondly, we do not consider the unbroken SM in a \emph{general Lorentz} gauge,
	in which case we are forced to take into account vertex renormalization,
	but choose to work within the unbroken SM 
	in a \emph{general background-field} gauge. 
	We keep the full dependence of the diagrams on the electroweak gauge-fixing parameters and
	take into account corresponding renormalization. Absence of these auxiliary parameters in the final expressions 
	for beta-functions gives us an independent confirmation of the correctness of our calculation.

	It is worth mentioning that in Refs.~\cite{Mihaila:2012fm,Mihaila:2012pz} the SM in BFG was also considered. 
	However, the corresponding calculation was carried out in the spontaneously broken phase and  
	the model file distributed with \texttt{FeynArts} package was used. 
	Since a consistent renormalization of the electroweak gauge-fixing parameters in 
	the spontaneously broken phase requires a severe modification
	of corresponding part of the model file (see, e.g., Refs.~\cite{Actis:2006ra,Actis:2006rb,Actis:2006rc}), 
	the Landau gauge was chosen in \cite{Mihaila:2012pz} to avoid these kind of problems.

And lastly, since the unbroken SM in BFG is not implemented as a \texttt{FeynArts} model file,
we are forced to use a package like \texttt{FeynRules} \cite{Christensen:2008py} or \texttt{LanHEP} \cite{Semenov:2010qt}.
Due to the fact that the authors are more accustomed to the latter, \texttt{LanHep} was chosen
to generate the required Feynman rules from the Lagrangian\footnote{The authors of Refs.~\cite{Mihaila:2012fm,Mihaila:2012pz} utilize \texttt{FeynRules} to obtain a model file for the unbroken SM.}.

The paper is organized as follows.
In Section~\ref{sec:unbroken_sm} we introduce our notation and
present a brief description of the unbroken SM quantized in the background-field gauge.
Section~\ref{sec:calc_details} describes the details of our calculation strategy.
Finally, the results and conclusions can be found in Section~\ref{sec:results}.
Appendix contains all the expressions for the considered renormalization constants.

\section{The Standard Model in the unbroken phase. The background-field gauge\label{sec:unbroken_sm}}

Let us briefly review the Lagrangian of the SM in the background-field gauge.
We closely follow \cite{Denner:1994xt} albeit the fact that we introduce background fields only for gauge bosons.
Moreover, as it was mentioned in Introduction, we neglect all the dimensionful couplings (i.e., mass parameters).

In our calculation we use the Lagrangian of the form 
\begin{equation}
{\mathcal L} =
  \LG 
+ \LH 
+ \LF 
+ \LGF 
+ \LFP. 
\label{eq:lag_classical}
\end{equation}
Here $\LG$ is the Yang-Mills part
\begin{eqnarray}
	\LG & = &
		-\frac{1}{4} G^a_{\mu\nu} G^a_{\mu\nu}
	        -\frac{1}{4} W^i_{\mu\nu} W^i_{\mu\nu}
	        -\frac{1}{4} B_{\mu\nu} B_{\mu\nu},
	\label{eq:lag_YM}\\
	G^a_{\mu\nu} & = & \partial_\mu G^a_\nu - \partial_\nu G^a_\mu + g_s f^{abc} G_\mu^b G_\nu^c,
	\label{eq:Gluon_fs} \\
	W^i_{\mu\nu} & = & \partial_\mu W^i_\nu - \partial_\nu W^i_\mu + g_2 \epsilon^{ijk} W^j_\mu W^k_\nu,
	\label{eq:Wboson_fs} \\
	B_{\mu\nu} & = & \partial_\mu B_\nu - \partial_\nu B_\mu,
	\label{eq:Bboson_fs}
\end{eqnarray}
where $G^a_\mu = \qG^a_\mu + \bgG^a_\mu$ ($a=1,\dots,8$), $W^i_\mu = \qW^i_\mu + \bgW^i_\mu $, ($i=1,2,3$), and $B_\mu = \qB_\mu + \bgB_\mu$ are
gauge fields for SU(3), SU(2) and U(1) groups. By $\qV=(\qG,\qW,\qB)$ we denote quantum fields, and $\bgV=(\bgG,\bgW,\bgB)$ is
used for their background counterpart.
The corresponding gauge couplings are $g_s$, $g_2$, and $g_1$.
The group structure constants enter into the commutation relations
\begin{equation}
	\left[ T^a, T^b \right]  =  i f^{abc} T^c,\qquad \left[ \tau^i, \tau^j \right]  =  i \epsilon^{ijk} \tau^k,
	\label{eq:SU3_SU2_str_const}
\end{equation}
	with $T^a = \lambda^a/2$ and $\tau^i = \sigma^i/2$ being color and weak isospin generators.

The covariant derivative acting on a field which is charged under all the gauge groups looks like
\begin{equation}
	D_\mu = \partial_\mu - i g_s T^a G^a_\mu - i g_2 \tau^i W^i_\mu + i g_1 \frac{Y_W}{2} B_\mu.
	\label{eq:cov_der}
\end{equation}
If a field is not charged under either group, the  corresponding term is omitted.
With the help of the covariant derivative one can write the following Higgs and fermionic parts of the Lagrangian:
\begin{eqnarray}
	\LH & = & \left( D_\mu \Phi \right)^\dagger  \left( D_\mu \Phi \right)
	- \lambda \left( \Phi^\dagger \Phi \right)^2,
	\label{eq:higgs_lag}\\
	\LF & = &
	\sum\limits_{i=1,2,3}
	    \bigg( i \bar{Q}^L_i  \Dhat Q^L_i  + i \bar L^L_i  \Dhat  L^L_i 	  
+  i \bar{u}^R_g  \Dhat  u^R_g +  i \bar{d}^R_g  \Dhat  d^R_g +  i \bar l^R_g  \hat D  l^R_g  \bigg)\nonumber\\
	& - & \sum\limits_{i,j=1,2,3} \bigg(
	Y^{ij}_u (Q^L_i \Phi^c) u^R_j
	+ Y^{ij}_d (Q^L_i \Phi) d^R_j  + Y^{ij}_l (L^L_i \Phi) l^R_j
	+ \mathrm{h.c.} \bigg),
	\label{eq:ferm_lag}
\end{eqnarray}
	where indices $i,j=1,2,3$ count different fermion families, $\lambda$ and $Y_{u,d,l}$ are the higgs quartic
	and  Yukawa matrices\footnote{In the actual calculation the diagonal Yukawa matrices were used. However, the result
	can be generalized with the help of additional tricks (see Sec.\ref{sec:calc_details} and Ref.~\cite{Mihaila:2012pz}).}, respectively.
The left-handed quarks $Q_g^L = (u_g, d_g)^L$ and leptons $L_g^L=(\nu_g,l_g)^L$ form the SU(2) doublets while
	the right-handed quarks ($u^R_g, d^R_g$) and charged leptons $l^R_g$ are the singlets with respect to SU(2).
The Higgs doublet $\Phi$ with $Y_W = 1$ has the following decomposition in terms of the component fields:
\begin{equation}
	\Phi =
	\left(
	\begin{array}{c}
		\phi^+(x) \\ \frac{1}{\sqrt 2} \left( h + i \chi \right)
		\end{array}
	\right),
	\qquad
	\Phi^c = i\sigma^2 \Phi^\dagger =
	\left(
	\begin{array}{c}
		\frac{1}{\sqrt 2} \left( h - i \chi \right) \\
		-\phi^-
		\end{array}
	\right).
	\label{eq:Phi_def}
\end{equation}
Here a charge-conjugated Higgs doublet is introduced $\Phi^c$ with $Y_W=-1$.

The gauge-fixing terms are introduced only for quantum fields
\begin{equation}
	\LGF  =
	-\frac{1}{2\xiG} G_G^a G_G^a
	-\frac{1}{2\xiW} G_W^i G_W^i
	-\frac{1}{2\xiB} G_B^2,
	\label{eq:gauge_fix_lag}
\end{equation}
with
\begin{eqnarray}
	G^a_G  & = & \partial_\mu \qG^a_\mu + g_s f^{abc} \bgG^b_\mu \qG^c_\mu\,, \nonumber\\ 
	G^i_W  & = & \partial_\mu \qW^i_\mu + g_2 \epsilon^{ijk} \bgW^j_\mu \qW^k_\mu\,, \nonumber\\ 
	G_B & = &  \partial_\mu \qB_\mu \, .  \label{eq:gauge_fix_func}
\end{eqnarray}
The ordinary derivatives are replaced by covariant ones containing the background fields.
Due to this, the invariance of the effective action under background gauge transformations is not touched by introduction of \eqref{eq:gauge_fix_lag}.

The Fadeev-Popov part of the Lagrangian is given by
\begin{equation}
	\LFP  =  - \bar c_\alpha \frac{\delta G_\alpha}{\delta \theta^\beta} c_\beta
	\label{eq:ghost_lag}
\end{equation}
where $\alpha,\beta = (G,W,B)$, and $\delta G_\alpha/\delta \theta^\beta$ is the variation of gauge-fixing functions \eqref{eq:gauge_fix_func}
under the following infinitesimal quantum gauge transformations
\begin{eqnarray}
	\delta \qG^a_\mu & = & (D_\mu \theta_G)^a = \partial_\mu \theta_G^a + g_s f^{abc} G^b_\mu \theta_G^c\,,\nonumber\\
	\delta \qW^i_\mu & = & (D_\mu \theta_W)^i = \partial_\mu \theta_W^i + g_2 \epsilon^{ijk} W^j_\mu \theta_W^k\,,\nonumber\\
	\delta \qB_\mu & = & \partial_\mu \theta_B\,.
	\label{eq:quantum_gauge_transform}
\end{eqnarray}
It should be stressed that covariant derivatives in \eqref{eq:quantum_gauge_transform} involve the sum of quantum and background gauge fields $V = \qV + \bgV$.
The corresponding background transformations are obtained from \eqref{eq:quantum_gauge_transform}
by the replacement $V\to\bgV$.

The Feynman rules for the model described by the Lagrangian \eqref{eq:lag_classical}  were generated
with the help of \texttt{LanHEP}  \footnote{\texttt{LanHEP} 3.1.5, which
was used by the authors, produces a wrong sign for the combination $f^{abc} f^{dec}$ during export to the \texttt{FeynArts} model files.
A new version with a fix is scheduled for November 2012.} \cite{Semenov:2010qt}.

It is worth mentioning here that our problem does not require the introduction of U(1) ghosts $\bar c_B, c_B$
	and background $\bgB$ fields.
This is due to the fact that the latter has the same interactions
	as its quantum counterpart $\qB$ and the former decouples from other particles.
Nevertheless, we keep them in our \texttt{LanHEP} model file to allow for possible generalizations
to non-linear gauge-fixing as in Ref.~\cite{Denner:1994xt}.

\section{Details of calculations \label{sec:calc_details}}

Due to the gauge invariance of the effective action for the background fields, QED-like Ward identities can be derived.
The latter can be used to prove the following simple relations:
\begin{equation}
	Z_{g_i} = Z_{\bgV_i}^{-1/2}, \qquad i=1,2,3
	\label{eq:bfm_rel}
\end{equation}
with $Z_{\bgV_i}$ and $Z_{g_i}$  being renormalization constants for background fields $\bgV^\mu_i = (\bgB^\mu, \bgW^\mu, \bgG^\mu)$ and
	SM gauge couplings  $g_i = (g_1,g_2, g_s)$, respectively.

Since we keep the full dependence on the gauge-fixing parameters
	$\xi_i $
	during the whole calculation,
        we also need to know how $\xi_i = (\xiB, \xiW, \xiG )$ are renormalized.
Again,  due to the Ward identities, the longitudinal part of the quantum gauge
	field propagators does not receive any loop corrections.
As a consequence, the following identities hold:
\begin{equation}
	Z_{\xi_i} = Z_{\qV_i}.	
	\label{eq:gauge_par_ren}
\end{equation}
Here $Z_{\xi_i}$ stands for the renormalization constants for the gauge-fixing parameters.
The quantum gauge fields $\qV_i$ are renormalized in the \MS-scheme with the help of $Z_{\qV_i}$.
It is clear from \eqref{eq:bfm_rel} and \eqref{eq:gauge_par_ren} that to carry out the calculation,
	one needs to consider gauge boson self-energies for both quantum $\qV$ and background $\bgV$ fields.

 For calculation of the renormalization constants, following~\cite{Larin:1993tp} (see also \cite{Tarasov:1976ef,Vladimirov:1979zm,Tarasov:1980kx}), we use
 the multiplicative renormalizability of the corresponding Green functions.
 The re\-normalization constants $Z_{V}$  relate the dimensionally
 regularized one-particle-irreducible two-point functions $\Gamma_{V,\mathrm{Bare}}$  
 with the renormalized one $\Gamma_{V,\mathrm{Ren}}$ as:
  \begin{equation}\label{multren}
   \Gamma_{V,\mathrm{Ren}}\left(\frac{Q^2}{\mu^2},a_i\right)=\lim_{\ep \rightarrow 0}
   Z_{V} \left(\frac{1}{\ep},a_i\right)
   \Gamma_{V,\mathrm{Bare}}\left(Q^2,a_{i,\mathrm{Bare}},\ep\right),
  \end{equation}
  where $a_{i,\mathrm{Bare}}$ are the bare parameters of the model.
  For convenience, we introduce the following notation, which is closely related to that used in Ref.~\cite{Mihaila:2012pz},
 \begin{eqnarray}
	 a_i  & = &  \left(\frac{5}{3} \frac{g_1^2}{16\pi^2}, \frac{g_2^2}{16\pi^2}, \frac{g_s^2}{16\pi^2}, \frac{Y_u^2}{16\pi^2}, \frac{Y_d^2}{16\pi^2}, \frac{Y_l^2}{16\pi^2}, \frac{\lambda}{16\pi^2},\xiG, \xiW, \xiG \right),
	 \label{eq:coupl_notations}
 \end{eqnarray}
	so we treat the gauge-fixing parameters along the same lines as couplings.
Moreover, in the renormalization group analysis of the SM
one usually employs the SU(5) normalization of the U(1) gauge coupling which leads to an additional factor $5/3$ in \eqref{eq:coupl_notations}.

	The bare parameters are related to the renormalized ones in the \MS-scheme by the following formula:
\begin{equation}
	a_{k,\mathrm{Bare}}\mu^{-2\rho_k\epsilon} = Z_{a_k} a_k(\mu)=a_k+\sum_{n=1}^\infty c_k^{(n)}\frac{1}{\epsilon^n}\,,
	\label{eq:bare_to_ren}
\end{equation}
where $\rho_k=1/2$ for the gauge ($g_1,g_2,~g_s$) and Yukawa constants ($Y_u,Y_d,~Y_l$), $\rho_k=1$ for the scalar quartic coupling constant $\lambda$, and $\rho_k=0$ for the gauge fixing parameters.
	In order to extract a three-loop contribution to $Z_V$ from the corresponding self-energies,
	it is sufficient to know the two-loop renormalization constants for the
	gauge couplings and the one-loop results for the Yukawa couplings.
	This is due to the fact that the Yukawa vertices appear for the first time only in the two-loop self-energies
	and the higgs self-coupling enters into the result only at the third level of perturbation theory.

The four-dimensional beta-functions, denoted by $\beta_i$, are defined via

\begin{equation}
		\beta_i(a_k) = \frac{d a_i(\mu,\epsilon)}{d \ln \mu^2}\bigg|_{\epsilon=0}\ .		
		\label{eq:beta_def}
\end{equation}
	Here, again, $a_i$ stands for both the gauge couplings and the gauge-fixing.

Given the fact that the bare parameters do not depend on the renormalization scale
	the expressions for $\beta_i$ can be obtained~\cite{Machacek:1983tz} by differentiation of \eqref{eq:bare_to_ren}
	with respect to $\ln \mu^2$:
	\begin{equation}
		-\rho_k\epsilon\bigg[a_k+\sum_{n=1}^\infty c_k^{(n)}\frac{1}{\epsilon^n}\bigg]=-\rho_k\epsilon a_k+\beta_k+\sum_{n=1}^\infty\sum_{l}(\beta_l-\rho_l a_l\epsilon)\frac{\partial c_k^{(n)}}{\partial a_l}\frac{1}{\epsilon^n}\,.
	\label{eq:beta_calc}
 	\end{equation}
 	Taking in account only the leading order of the expansion in $\epsilon$:
	\begin{equation}
 	\beta_k =  \sum_{l}\rho_l a_l\frac{\partial c_k^{(1)}}{\partial a_l}-\rho_k c_k^{(1)}\,.
	 	\label{eq:beta_calc_simple}
	\end{equation}

In \MS-like schemes the renormalization constants for the Green functions may be expanded as
\begin{equation}
	Z_{\Gamma} = 1 + \sum\limits_{k=1}^{\infty} \frac{Z_{\Gamma}^{(k)}}{\ep^k}\,.
	\label{eq:ren_const_ep_decom}
\end{equation}
Differentiating \eqref{eq:ren_const_ep_decom} with respect to $\ln \mu^2$ we simply get all-order
expression for anomalous dimensions:
\begin{equation}
	\gamma_\Gamma\equiv - \mu^2 \frac{\partial \ln Z_\Gamma}{\partial \mu^2}
	=  - \left[\sum\limits_{j} \big(\beta_j -\rho_j a_j\epsilon\big)\frac{\partial Z_\Gamma}{\partial a_j}\right] Z_\Gamma^{-1}\,.
	\label{eq:gamma_all_ord}
\end{equation}

It turns out that the above expression is finite as $\ep\to0$ so 
\begin{equation}
	\gamma_\Gamma= \sum\limits_j a_j \rho_j \frac{\partial Z^{(1)}_\Gamma}{\partial a_j}\,.
	\label{eq:gamma_zero_ord}
\end{equation}
The advantage of \eqref{eq:beta_calc} and \eqref{eq:gamma_all_ord} comes from the fact that
	it provides us with additional confirmation of the correctness of the final result since
	beta functions and anomalous dimensions extracted directly from \eqref{eq:beta_calc} and \eqref{eq:gamma_all_ord}
	are finite for $\ep \to 0$ only if $c_{k}^{(n)}$ satisfy
	the so-called pole equations \cite{'tHooft:1972fi}, e.g.,
\begin{equation}
	\left[\sum\limits_l \rho_l a_l \frac{\partial }{\partial a_l}
	- \rho_k \right] c_k^{(n+1)}
	= \sum_{l}\beta_l\frac{\partial c_k^{(n)}}{\partial a_l}\,.
	\label{eq:pole_eq}
\end{equation}

In order to calculate  the bare two-point functions for the quantum and background fields, we generate
	the corresponding diagrams with the help of the \texttt{FeynArts} package~\cite{Hahn:2000kx}.
It is worth pointing that we use the \texttt{Classes} level of diagram generation which allows us to significantly
	reduce the number of generated diagrams since we do not distinguish fermion generations.
The complexity of the problem can be deduced from Table \ref{tab:dia_num}
	that shows how the number of the \texttt{FeynArts} generated diagrams increases with the loop level.
Clearly, the presented numbers are an order of magnitude less than those given in Table I of
	Ref.~\cite{Mihaila:2012pz}, which somehow demonstrate the advantage of our approach.

The number of the SM fermion generations is introduced via counting fermion traces present
in the generated expression for a diagram and multiplying it by $\NGen$.
We separately count fermion traces involving the Yukawa interaction vertices and multiply them not by $\NGen$ but by $\NY$.
This allows us to use the following substitution rules (c.f.,~\cite{Mihaila:2012pz}) to generalize the obtained
	expression to the case of the general Yukawa matrices
\begin{eqnarray}
  \NY \big[ a_u, a_d, a_l \big] & \rightarrow & \big[ \YtU, \YtD, \YtL \big],\nonumber\\
  \NY \big[ a^2_u, a^2_d, a^2_l \big] & \rightarrow & \big[ \YtUs, \YtDs, \YtLs \big],\nonumber\\
  \NY^2 \big[ a^2_u, a^2_d, a^2_l \big] & \rightarrow & \big[ \YtUU, \YtDD, \YtLL \big],\nonumber\\
  \NY^2 \big[ a_u a_d, a_d a_l, a_u a_l \big] & \rightarrow & \big[ \YtU \YtD, \YtD \YtL, \YtU \YtL \big],\nonumber\\
  \NY a_u a_d & \rightarrow & \YtUD
\label{eq:genr_yuk_tr}
\end{eqnarray}
	where
\begin{equation}
	  \YtU =  \frac{\tr Y_u Y_u^\dagger}{16\pi^2},
	\qquad \YtD =  \frac{\tr Y_d Y_d^\dagger}{16\pi^2},
	\qquad \YtL =  \frac{\tr Y_l Y_l^\dagger}{16\pi^2},
	\label{eq:yukawa_tr_l_1}
\end{equation}
	and
\begin{eqnarray}
	  \YtUs =  \frac{\tr Y_u Y_u^\dagger Y_u Y_u^\dagger}{(16\pi^2)^2},
	& \qquad &
	  \YtDs =  \frac{\tr Y_d Y_d^\dagger Y_d Y_d^\dagger}{(16\pi^2)^2}, \nonumber\\
	  \YtUD =  \frac{\tr Y_u Y_u^\dagger Y_d Y_d^\dagger}{(16\pi^2)^2},
	& \qquad &
	  \YtLs =  \frac{\tr Y_l Y_l^\dagger Y_l Y_l^\dagger}{(16\pi^2)^2}.
	\label{eq:yukawa_tr_l_2}
\end{eqnarray}

A comment is in order about the last substitution in \eqref{eq:genr_yuk_tr}.
It turns out that $\YtUD$ is the only combination of up- and down-type Yukawa matrices,
which can appear in the result for the three-loop gauge-boson self-energy within the SM.
This can be traced to the following facts:
1) in the unbroken SM all the particles are massless so that chirality is conserved 
	during fermion propagation;
2) only the Yukawa interactions flip the chirality of the incoming fermions;
3) there is no right-handed flavour changing current coupled to a SM gauge field.
As a consequence, combinations like
\begin{equation}
	\frac{\tr Y_u Y_d^\dagger Y_u Y_d^\dagger}{(16\pi^2)^2} \qquad \mbox{and} \qquad
	\frac{\tr Y_u Y_d^\dagger Y_d Y_u^\dagger}{(16\pi^2)^2},
\end{equation}
	which require at least two chirality-conserving transitions between right-handed up- and down-type quarks,
	do not show up in the result.

	This type of counting is performed at the generation stage.
A simple script converts the output of \texttt{FeynArts} to \texttt{DIANA}-like \cite{Tentyukov:1999is} notation and identifies
	\texttt{MINCER} topologies.
This allows us to use the \texttt{FORM} \cite{Vermaseren:2000nd} package
	\texttt{COLOR} \cite{vanRitbergen:1998pn} to do the SU(3) color algebra and \texttt{MINCER}
	\cite{Gorishnii:1989gt} to obtain the \ep-expansion of diagrams.
It is worth pointing that the expressions for all SM gauge couplings exhibit explicit dependence on number of
	colors $\NR$ which stems from the fact that we have to sum over color when
	there is a (sub)loop with external color singlets coupled to quarks.

During our calculation we made use of naive anticommuting prescription
	for dealing with $\gamma_5$ (see, e.g., a nice review \cite{Jegerlehner:2000dz}).
	In this case, however, closed fermion loops with odd number of $\gamma_5$ (``odd traces'') are not treated properly.
	In $D=4$ such traces inevitably lead to the appearance of
	four-dimensional antisymmetric tensors $\eps_{\mu\nu\rho\sigma}$ that in the final result 
	for a diagram should be contracted either between themselves or with external Lorentz indices and/or 
	momenta. 
	Since we are only interested in two-point functions the only non-zero combination 
	that could potentially appear after loop integration is 
	$\eps_{\mu\alpha\rho\sigma} \eps_{\nu\beta\rho\sigma} q_\alpha q_\beta$ which originates
	at three loops from two odd traces.  
	In this expression $q$ corresponds to external momentum, $\mu$, $\nu$ denote 
	external Lorentz indices, and $\rho$, $\sigma$ are dummy indices representing the contractions
	due to internal vector boson propagators.
	A simple counting shows that both closed fermion lines are one-loop triangle (sub)graphs 
	contributing to Adler-Bell-Jackiw
	gauge anomalies \cite{Adler:1969gk,Bell:1969ts,Bardeen:1969md}.
	Having in mind the cancellation of such anomalies within the SM \cite{Bouchiat:1972iq,Gross:1972pv}, 
	in our calculation we can safely put all the Dirac traces involving odd number of $\gamma_5$ to zero  
	(see also the discussion in Ref.~\cite{Mihaila:2012pz}).
	It is worth mentioning that the correct results for three-loop contribution to Yukawa coupling beta-functions 
	\cite{Chetyrkin:2012rz}	can not be obtained without special treatment of such traces.

\begin{table}
\begin{center}
    \begin{tabular}{| l || c | c | c || c || c | c | c |}
    \hline
    Broken        & 1    & 2 & 3 & Unbroken & 1 & 2 & 3 \\ \hline
    $W^{+}/W^{-}$ & 10   & 339 & 21942 &   $\bgW^i$      & 11 & 389 & 36647  \\ \hline
    $Z$           & 9    & 281 & 19041 &   $\qW^i$      & 11 & 371 & 36103 \\ \hline
    $A$           & 7    & 218 & 14426 &   $\bgB,\qB$      & 6 & 214 & 20144 \\ \hline
    $ZA$          & 7    & 236 & 16120 &   $\bgG$     & 4 & 73 & 4183  \\ \hline
    $G$           & 4    & 67 & 3287   &   $\qG$  & 4 & 66 & 4060    \\ \hline
    Total       & 37    & 1141 & 74816 &   Total      & 36 & 1113 & 101137 \\
    \hline
  \end{tabular}
  \caption{Number of self-energy diagrams with external gauge fields,
  	generated by \texttt{FeynArts} in the broken and unbroken SM,
  	at one, two, and three loops. }
  \label{tab:dia_num}
  \end{center}
\end{table}

\section{Results and conclusions \label{sec:results}}

Here we present the results of our calculations in the form of the SM gauge beta-functions and anomalous
	dimension of the gauge-fixing parameters.
From \eqref{eq:bfm_rel} and \eqref{eq:gauge_par_ren} it is clear that anomalous dimensions of
	the background fields are connected with the corresponding gauge coupling beta-functions
\begin{equation}
	\gamma_{\bgB} = - \beta_1/a_1,
	\qquad
	\gamma_{\bgW} = - \beta_2/a_2,
	\qquad
	\gamma_{\bgG} = - \beta_s/a_s
	\label{eq:bgf_anom_dim}
\end{equation}
	and for the quantum fields we have
\begin{equation}
	\gamma_{\qB} = \beta_{\xiB}/\xiB,
	\qquad
	\gamma_{\qW} = \beta_{\xiW}/\xiW,	
	\qquad
	\gamma_{\qG} = \beta_{\xiG}/\xiG.	
	\label{eq:qf_anom_dim}
\end{equation}
	The corresponding renormalization constants can be found in the Appendix.

	At the end of the day, we have the following expressions for the beta-functions ($\lam\equiv a_\lambda$):
{\allowdisplaybreaks
\begin{align}
\beta_1 & =\ala^2 \bigg(\NGen \bigg(\frac{11 \
\NR}{45}+\frac{3}{5}\bigg)+\frac{1}{10}\bigg) \notag \\
 & + \ala^2 \bigg(\NGen \bigg(\frac{137 \ala \
\NR}{900}+\frac{81 \ala}{100}+\frac{\alb \
\NR}{20}+\frac{9 \alb}{20}+\frac{11 \alc \cR \
\NR}{15}\bigg) \notag \\
 & +\frac{9 \ala}{50}+\frac{9 \
\alb}{10} -\frac{\NR \YtD}{6}-\frac{17 \NR \
\YtU}{30}-\frac{3 \YtL}{2}\bigg)  \notag \\
 & + \ala^2 \bigg(\NGen \bigg(-\frac{1697 \ala^2 \
\NR}{18000}-\frac{981 \ala^2}{2000}-\frac{\ala \alb \
\NR}{1200}-\frac{27 \ala \alb}{400} \notag \\
 &-\frac{137}{900} \
\ala \alc \cR \NR+\frac{\alb^2 \
\NR}{45}+\frac{27 \alb^2}{10}-\frac{1}{20} \alb \alc \
\cR \NR+\frac{1463}{540} \alc^2 \cA \cR \
\NR \notag \\
 & -\frac{11}{30} \alc^2 \cR^2 \NR\bigg)+\NGen^2 \
\bigg(-\frac{16577 \ala^2 \NR^2}{486000}-\frac{2387 \ala^2 \
\NR}{9000}-\frac{891 \ala^2}{2000}-\frac{11 \alb^2 \
\NR^2}{720} \notag \\
 & -\frac{11 \alb^2 \NR}{72}-\frac{11 \
\alb^2}{80}-\frac{242}{135} \alc^2 \cR \ItoR \
\NR\bigg)+\frac{489 \ala^2}{8000}+\frac{783 \ala \
\alb}{800}+\frac{27 \ala \lam}{50} \notag \\
 & -\frac{1267 \ala \
\NR \YtD}{2400}-\frac{2827 \ala \NR \
\YtU}{2400}-\frac{2529 \ala \YtL}{800}+\frac{3401 \
\alb^2}{320}+\frac{9 \alb \lam}{10} \notag \\
 &-\frac{437 \alb \
\NR \YtD}{160} -\frac{157 \alb \NR \
\YtU}{32}-\frac{1629 \alb \YtL}{160}-\frac{17}{20} \
\alc \cR \NR \YtD-\frac{29}{20} \alc \cR \
\NR \YtU \notag \\
 & -\frac{9 \lam^2}{5}+\frac{17 \NR^2 \
\YtDD}{120}+\frac{59}{60} \NR^2 \YtD \YtU+\frac{101 \
\NR^2 \YtUU}{120} +\frac{157 \NR \YtD \
\YtL}{60}+\frac{61 \NR \YtDs}{80} \notag \\
 & +\frac{199 \NR \
\YtL \YtU}{60}+\frac{\NR \YtUD}{8}+\frac{113 \
\NR \YtUs}{80}+\frac{99 \YtLL}{40}+\frac{261 \
\YtLs}{80}\bigg)\,,
\end{align}

\begin{align}
\beta_2 & =
\alb^2 \bigg(\NGen \
\bigg(\frac{\NR}{3}+\frac{1}{3}\bigg)-\frac{43}{6}\bigg) \notag \\
 & + \alb^2 \bigg(\NGen \bigg(\frac{\ala \NR}{60}+\frac{3 \
\ala}{20}+\frac{49 \alb \NR}{12}+\frac{49 \
\alb}{12}+\alc \cR \NR\bigg) \notag \\
 & +\frac{3 \
\ala}{10}-\frac{259 \alb}{6}-\frac{\NR \
\YtD}{2}-\frac{\NR \YtU}{2}-\frac{\YtL}{2}\bigg) \notag \\
 & + \alb^2 \bigg(\NGen \bigg(-\frac{287 \ala^2 \
\NR}{3600}-\frac{91 \ala^2}{400}+\frac{13 \ala \alb \
\NR}{240}+\frac{39 \ala \alb}{80}-\frac{1}{60} \ala \
\alc \cR \NR \notag \\
 & +\frac{1603 \alb^2 \
\NR}{27}+\frac{1603 \alb^2}{27}+\frac{13}{4} \alb \
\alc \cR \NR+\frac{133}{36} \alc^2 \cA \
\cR \NR-\frac{1}{2} \alc^2 \cR^2 \
\NR\bigg) \notag \\
 & +\NGen^2 \bigg(-\frac{121 \ala^2 \
\NR^2}{32400}-\frac{77 \ala^2 \NR}{1800}-\frac{33 \
\ala^2}{400}-\frac{415 \alb^2 \NR^2}{432}-\frac{415 \
\alb^2 \NR}{216}-\frac{415 \alb^2}{432} \notag \\
 & -\frac{22}{9} \
\alc^2 \cR \ItoR \NR\bigg)+\frac{163 \
\ala^2}{1600}+\frac{561 \ala \alb}{160}+\frac{3 \ala \
\lam}{10}-\frac{533 \ala \NR \YtD}{480}-\frac{593 \
\ala \NR \YtU}{480} \notag \\
 & -\frac{51 \ala \
\YtL}{32}-\frac{667111 \alb^2}{1728}+\frac{3 \alb \
\lam}{2}-\frac{243 \alb \NR \YtD}{32}-\frac{243 \
\alb \NR \YtU}{32}-\frac{243 \alb \
\YtL}{32} \notag \\
 & -\frac{7}{4} \alc \cR \NR \
\YtD-\frac{7}{4} \alc \cR \NR \YtU-3 \
\lam^2+\frac{5 \NR^2 \YtDD}{8}+\frac{5}{4} \NR^2 \
\YtD \YtU+\frac{5 \NR^2 \YtUU}{8} \notag \\
 & +\frac{5 \NR \
\YtD \YtL}{4}+\frac{19 \NR \YtDs}{16}+\frac{5 \
\NR \YtL \YtU}{4}+\frac{9 \NR \
\YtUD}{8}+\frac{19 \NR \YtUs}{16} \notag \\
 & +\frac{5\YtLL}{8}+\frac{19 \YtLs}{16}\bigg)\,,
\end{align}


\begin{align}
\beta_s & =
\alc^2 \bigg(\frac{8 \ItoR \NGen}{3}-\frac{11 \
\cA}{3}\bigg) \notag \\
 & + \alc^2 \bigg(\NGen \bigg(\frac{11 \ala \ItoR}{15}+3 \
\alb \ItoR+\frac{40 \alc \cA \ItoR}{3}+8 \
\alc \cR \ItoR\bigg) \notag \\
 & -\frac{34 \alc \cA^2}{3}-4 \
\ItoR \YtD-4 \ItoR \YtU\bigg)  + \alc^2 \bigg(\NGen \bigg(-\frac{13 \ala^2 \
\ItoR}{60}-\frac{\ala \alb \ItoR}{20}\notag \\
 &+\frac{22}{15} \
\ala \alc \cA \ItoR-\frac{11}{15} \ala \
\alc \cR \ItoR+\frac{241 \alb^2 \ItoR}{12} \notag\\
& +6\alb \alc \cA \ItoR-3 \alb \alc \cR \
\ItoR+\frac{2830}{27} \alc^2 \cA^2 \ItoR+\frac{410}{9} \
\alc^2 \cA \cR \ItoR-4 \alc^2 \cR^2 \
\ItoR\bigg) \notag \\
 & +\NGen^2 \bigg(-\frac{1331 \ala^2 \ItoR \
\NR}{8100}-\frac{121 \ala^2 \ItoR}{300}-\frac{11}{12} \
\alb^2 \ItoR \NR-\frac{11 \alb^2 \
\ItoR}{12}-\frac{632}{27} \alc^2 \cA \ItoR^2 \notag \\
 & -\frac{176}{9} \alc^2 \cR \ItoR^2\bigg)-\frac{89 \ala \ItoR \YtD}{20}-\frac{101 \
\ala \ItoR \YtU}{20}-\frac{93 \alb \ItoR \
\YtD}{4}-\frac{93 \alb \ItoR \YtU}{4} \notag \\
 & -\frac{2857}{54} \
\alc^2 \cA^3 -24 \alc \cA \ItoR \YtD-24 \
\alc \cA \ItoR \YtU-6 \alc \cR \ItoR \
\YtD-6 \alc \cR \ItoR \YtU \notag \\
 & +7 \ItoR \
\NR \YtDD+14 \ItoR \NR \YtD \YtU+7 \
\ItoR \NR \YtUU+7 \ItoR \YtD \YtL+9 \
\ItoR \YtDs+7 \ItoR \YtL \YtU \notag \\
 & -6 \ItoR \
\YtUD+9 \ItoR \YtUs\bigg)\,,
\end{align}
\begin{align}
\beta_{\xi_B} & =\xiB \bigg(-\frac{\ala}{10}+\NGen \bigg(-\frac{3 \ala}{5}-\frac{11 \ala \NR}{45}\bigg)\bigg) \notag \\
 & + \xiB \bigg(-\frac{9 \ala^2}{50}-\frac{9 \ala \alb}{10}+\frac{\ala \NR \YtD}{6}+\frac{3 \ala \YtL}{2}+\frac{17 \ala \NR \YtU}{30} \notag \\
 & +\NGen
\bigg(-\frac{81 \ala^2}{100}-\frac{9 \ala \alb}{20}-\frac{137 \ala^2
  \NR}{900}-\frac{\ala \alb \NR}{20}-\frac{11}{15} \ala \alc \cR\NR\bigg)\bigg) \notag \\
 & + \xiB \bigg(-\frac{489 \ala^3}{8000}-\frac{783 \ala^2
  \alb}{800}-\frac{3401 \ala \alb^2}{320}-\frac{27 \ala^2
  \lam}{50}-\frac{9 \ala \alb \lam}{10}+\frac{9 \ala \lam^2}{5} \notag \\
 & +\NGen \bigg(\frac{981 \ala^3}{2000}+\frac{27 \ala^2
  \alb}{400}-\frac{27 \ala \alb^2}{10}+\frac{1697 \ala^3
  \NR}{18000}+\frac{\ala^2 \alb \NR}{1200}-\frac{1}{45} \ala \alb^2
\NR \notag \\
& +\frac{137}{900} \ala^2 \alc \cR \NR+\frac{1}{20} \ala \alb \alc \cR \NR-\frac{1463}{540} \ala \alc^2
\cA \cR \NR+\frac{11}{30} \ala \alc^2 \cR^2 \NR\bigg) \notag \\
 & +\NGen^2
\bigg(\frac{891 \ala^3}{2000}+\frac{11 \ala \alb^2}{80}+\frac{2387 \ala^3 \NR}{9000}+\frac{11}{72} \ala \alb^2
\NR+\frac{242}{135} \ala \alc^2 \cR \ItoR \NR \notag \\
 & +\frac{16577 \ala^3
  \NR^2}{486000}+\frac{11}{720} \ala \alb^2 \NR^2\bigg)+\frac{1267 \ala^2 \NR \YtD}{2400}+\frac{437}{160} \ala \alb \NR
\YtD \notag \\
 & +\frac{17}{20} \ala \alc \cR \NR \YtD-\frac{17}{120} \ala \NR^2
\YtDD-\frac{61 \ala \NR \YtDs}{80}+\frac{2529 \ala^2 \YtL}{800}+\frac{1629 \ala \alb
  \YtL}{160} \notag \\
 & -\frac{157}{60} \ala \NR \YtD \YtL-\frac{99 \ala
  \YtLL}{40}-\frac{261 \ala \YtLs}{80}+\frac{2827 \ala^2 \NR
  \YtU}{2400}+\frac{157}{32} \ala \alb \NR \YtU \notag \\
  & +\frac{29}{20} \ala \alc \cR \NR
\YtU-\frac{59}{60} \ala \NR^2 \YtD \YtU-\frac{199}{60} \ala \NR \YtL
\YtU-\frac{101}{120} \ala \NR^2 \YtUU \notag \\
 & -\frac{\ala \NR \YtUD}{8}-\frac{113 \ala \NR \YtUs}{80}\bigg)\,,
\end{align}


\begin{align}
\beta_{\xi_W} & =\xiW \bigg(\frac{25 \alb}{6}-\alb \xiW+\NGen \bigg(-\frac{\alb}{3}-\frac{\alb \NR}{3}\bigg)\bigg) \notag \\
 & + \xiW \bigg(-\frac{3 \ala \alb}{10}+\frac{113 \alb^2}{4}-\frac{11
  \alb^2 \xiW}{2}-\alb^2 \xiW^2+\NGen \bigg(-\frac{3 \ala \alb}{20} \notag \\
 & -\frac{13 \alb^2}{4}-\frac{\ala \alb \NR}{60}-\frac{13 \alb^2 \NR}{4}-\alb \alc \cR \NR\bigg)+\frac{\alb \NR \YtD}{2}+\frac{\alb \YtL}{2}+\frac{\alb \NR \YtU}{2}\bigg) \notag \\
 & + \xiW \bigg(-\frac{163 \ala^2 \alb}{1600}-\frac{33 \ala
  \alb^2}{32}+\frac{143537 \alb^3}{576}-\frac{3 \ala \alb \lam}{10}-
  \frac{3 \alb^2 \lam}{2}+3 \alb \lam^2 \notag \\
 & -\frac{315 \alb^3 \xiW}{8}-\frac{33 \alb^3 \xiW^2}{4}-\frac{7 \alb^3 \xiW^3}{4}+\NGen^2
\bigg(\frac{33 \ala^2 \alb}{400}+\frac{185 \alb^3}{144}+\frac{77
  \ala^2 \alb \NR}{1800} \notag \\
 & +\frac{185 \alb^3 \NR}{72}+\frac{22}{9} \alb \alc^2 \cR \ItoR \NR+\frac{121 \ala^2 \alb
  \NR^2}{32400}+\frac{185 \alb^3 \NR^2}{144}\bigg)+\frac{533}{480}
\ala \alb \NR \YtD \notag \\
 & +\frac{79}{32} \alb^2 \NR \YtD+\frac{7}{4} \alb \alc \cR \NR \YtD-\frac{5}{8} \alb \NR^2
\YtDD-\frac{19 \alb \NR \YtDs}{16}+\frac{51 \ala \alb
  \YtL}{32} \notag \\
 & +\frac{79 \alb^2 \YtL}{32}-\frac{5}{4} \alb \NR \YtD \YtL-\frac{5 \alb \YtLL}{8}-\frac{19
  \alb \YtLs}{16}+\frac{593}{480} \ala \alb \NR \YtU+\frac{79}{32}
\alb^2 \NR \YtU \notag \\
 & +\frac{7}{4} \alb \alc \cR \NR \YtU-\frac{5}{4} \alb \NR^2 \YtD
\YtU-\frac{5}{4} \alb \NR \YtL \YtU-\frac{5}{8} \alb \NR^2
\YtUU-\frac{9 \alb \NR \YtUD}{8} \notag \\
 & -\frac{19 \alb \NR \YtUs}{16}-\frac{9}{10} \ala \alb^2
\zetathree+\frac{3 \alb^3 \zetathree}{2}-6 \alb^3 \xiW
\zetathree-\frac{3}{2} \alb^3 \xiW^2 \zetathree \notag \\
 & +\NGen \bigg(\frac{91
  \ala^2 \alb}{400}+\frac{6 \ala \alb^2}{5}-\frac{7025 \alb^3}{144}+2 \alb^3
\xiW+\frac{287 \ala^2 \alb \NR}{3600}+\frac{2}{15} \ala \alb^2
\NR \notag \\
 & -\frac{7025 \alb^3 \NR}{144}+\frac{1}{60} \ala \alb \alc \cR \NR+8 \alb^2 \alc \cR \NR-\frac{133}{36} \alb \alc^2 \cA \cR
\NR \notag \\
 & +\frac{1}{2} \alb \alc^2 \cR^2 \NR+2 \alb^3 \xiW \NR-\frac{9}{5}
\ala \alb^2 \zetathree+9 \alb^3 \zetathree-\frac{1}{5} \ala \alb^2 \NR \zetathree \notag \\
 & +9 \alb^3 \NR \zetathree-12 \alb^2 \alc \cR \NR \zetathree\bigg)\bigg)\,,
\end{align}


\begin{align}
\beta_{\xi_G} & = \xiG \bigg(\frac{13 \alc \cA}{6}-\frac{\alc \cA \xiG}{2}-\frac{8 \alc \ItoR \NGen}{3}\bigg) \notag \\
 & + \xiG \bigg(\frac{59 \alc^2 \cA^2}{8}-\frac{11}{8} \alc^2 \cA^2
\xiG-\frac{1}{4} \alc^2 \cA^2 \xiG^2+4 \alc \ItoR \YtD+4 \alc \ItoR \YtU \notag \\
 & +\bigg(-\frac{11}{15} \ala \alc \ItoR-3 \alb \alc \ItoR-10 \alc^2 \cA \ItoR-8 \alc^2 \cR \ItoR\bigg) \NGen\bigg) \notag \\
 & + \xiG \bigg(\frac{9965 \alc^3 \cA^3}{288}-\frac{167}{32} \alc^3
\cA^3 \xiG-\frac{33}{32} \alc^3 \cA^3 \xiG^2-\frac{7}{32} \alc^3 \cA^3
\xiG^3+\NGen^2 \bigg(\frac{121}{300} \ala^2 \alc \ItoR \notag \\
 & +\frac{11}{12} \alb^2 \alc \ItoR+\frac{304}{9} \alc^3 \cA
\ItoR^2+\frac{176}{9} \alc^3 \cR \ItoR^2+\frac{1331 \ala^2 \alc \ItoR
  \NR}{8100}+\frac{11}{12} \alb^2 \alc \ItoR \NR\bigg) \notag \\
 & +\frac{89}{20} \ala \alc \ItoR \YtD+\frac{93}{4} \alb \alc \ItoR
\YtD+\frac{25}{2} \alc^2 \cA \ItoR \YtD+6 \alc^2 \cR \ItoR \YtD-7 \alc
\ItoR \NR \YtDD \notag \\
 & -9 \alc \ItoR \YtDs-7 \alc \ItoR \YtD \YtL+\frac{101}{20} \ala \alc
\ItoR \YtU+\frac{93}{4} \alb \alc \ItoR \YtU+\frac{25}{2} \alc^2 \cA
\ItoR \YtU \notag \\
 & +6 \alc^2 \cR \ItoR \YtU-14 \alc \ItoR \NR \YtD \YtU-7 \alc \ItoR
\YtL \YtU-7 \alc \ItoR \NR \YtUU+6 \alc \ItoR \YtUD \notag \\
 & -9 \alc \ItoR \YtUs-\frac{9}{16} \alc^3 \cA^3
\zetathree-\frac{3}{4} \alc^3 \cA^3 \xiG \zetathree-\frac{3}{16}
\alc^3 \cA^3 \xiG^2 \zetathree+\NGen \bigg(\frac{13}{60} \ala^2 \alc
\ItoR \notag \\
 & +\frac{1}{20} \ala \alb \alc \ItoR-\frac{241}{12} \alb^2 \alc
\ItoR+\frac{319}{120} \ala \alc^2 \cA \ItoR+\frac{87}{8} \alb \alc^2
\cA \ItoR-\frac{911}{9} \alc^3 \cA^2 \ItoR \notag \\
 & +\frac{11}{15} \ala \alc^2 \cR \ItoR+3 \alb \alc^2 \cR
\ItoR-\frac{5}{9} \alc^3 \cA \cR \ItoR+4 \alc^3 \cR^2 \ItoR+4 \alc^3
\cA^2 \xiG \ItoR \notag \\
 & -\frac{22}{5} \ala \alc^2 \cA \ItoR \zetathree-18 \alb \alc^2 \cA \ItoR \zetathree+36 \alc^3 \cA^2 \ItoR \zetathree\notag \\
 & -48 \alc^3 \cA \cR \ItoR \zetathree\bigg)\bigg)\,.
\end{align}
}

With the help of substitutions $\cA = \NR= 3$, $\cR=4/3$, $\ItoR=1/2$,
	$\YtU = \tr \hat T$,
	$\YtD = \tr \hat B$,
	$\YtL = \tr \hat L$,
	$\YtDs = \tr (\hat B^2)$,
	$\YtUs = \tr (\hat T^2)$,
	$\YtLs = \tr (\hat L^2)$, and
	$\YtUD = \tr \hat T \hat B$
	it is possible to prove that the expressions presented above coincide with the results
	for the gauge beta functions
	obtained in Ref.~\cite{Mihaila:2012fm}.

As a consequence, one can be sure that the three-loop renormalization group equations obtained for the first time in Ref.~\cite{Mihaila:2012fm}
	are correct and confirmed by an independent calculation.
It is also worth mentioning that the obtained results can be used not only for the analysis of vacuum stability constraints within the SM
	(as in Refs.~\cite{Bezrukov:2012sa,Degrassi:2012ry,Alekhin:2012py})
but also, e.g.,  for very precise matching of the SM with its supersymmetric extension since the corresponding three-loop renormalization group functions
are already known from the literature \cite{Ferreira:1996ug,Jack:2004ch,Harlander:2009mn}.
Moreover, the leading two-loop decoupling corrections for the strongest SM couplings are also calculated within
the MSSM in Refs.~\cite{Bednyakov:2007vm,Harlander:2007wh,Bauer:2008bj,Bednyakov:2009wt}.

\subsection*{Acknowledgments}
The authors would like to thank M.~Kalmykov for drawing our attention to the problem and
A.~Semenov for his help with the \texttt{LanHEP} package.
This work is partially supported by RFBR grants 11-02-01177-a, 12-02-00412-a, RSGSS-4801.2012.2

\appendix
\section{Renormalization constants}
	Here we present the results for the renormalization constants from which the anomalous dimensions
	and beta-functions were extracted. It should be pointed out that the coefficients
	of the $\ep$-expansion satisfy the pole equations \eqref{eq:pole_eq}. The corresponding expressions
	together with the results for beta-functions can be found online\footnote{As ancillary files of the
	\texttt{arXiv} version of the paper} in the form of \texttt{Mathematica} files.
	
{\allowdisplaybreaks
\begin{align}
Z_{\alpha_1} & = 1 + \ala\invep\bigg\{\NGen \bigg(\frac{11 \NR}{45}+\frac{3}{5}\bigg)+\frac{1}{10}\bigg\} \notag \\
 & + \ala\bigg\{\epp\bigg[\NGen^2 \bigg(\frac{121 \ala \NR^2}{2025}+\frac{22 \ala \NR}{75}+\frac{9
   \ala}{25}\bigg)+\NGen \bigg(\frac{11 \ala \NR}{225}+\frac{3
   \ala}{25}\bigg)+\frac{\ala}{100} \bigg]  \notag \\
 & + \invep\bigg[ \NGen \bigg(\frac{137 \ala \NR}{1800}+\frac{81 \ala}{200}+\frac{\alb \NR}{40}+\frac{9
   \alb}{40}+\frac{11 \alc \cR \NR}{30}\bigg)+\frac{9 \ala}{100}+\frac{9
   \alb}{20} \notag \\
 & -\frac{\NR \YtD}{12}-\frac{17 \NR \YtU}{60}-\frac{3 \YtL}{4}\bigg]\bigg\} \notag \\
 & + \ala\bigg\{\eppp\bigg[\NGen^3 \bigg(\frac{1331 \ala^2 \NR^3}{91125}+\frac{121 \ala^2 \NR^2}{1125}+\frac{33 \ala^2
   \NR}{125}+\frac{27 \ala^2}{125}\bigg) \notag \\
 & +\NGen^2 \bigg(\frac{121 \ala^2 \NR^2}{6750}+\frac{11
   \ala^2 \NR}{125}+\frac{27 \ala^2}{250}\bigg)+\NGen \bigg(\frac{11 \ala^2 \NR}{1500}+\frac{9
   \ala^2}{500}\bigg)+\frac{\ala^2}{1000} \bigg]  \notag \\
 &+ \epp\bigg[ \NGen \bigg(\frac{3731 \ala^2 \NR}{54000}+\frac{441 \ala^2}{2000}+\frac{9 \ala \alb
   \NR}{40}+\frac{117 \ala \alb}{200}+\frac{11}{150} \ala \alc \cR \NR \notag \\
 & -\frac{11}{270} \ala \NR^2 \YtD-\frac{187 \ala \NR^2 \YtU}{1350}-\frac{\ala \NR
   \YtD}{10}-\frac{11 \ala \NR \YtL}{30}-\frac{17 \ala \NR \YtU}{50}-\frac{9 \ala
   \YtL}{10} \notag \\
 & -\frac{7 \alb^2 \NR}{720}-\frac{39 \alb^2}{80}-\frac{121}{270} \alc^2 \cA
   \cR \NR\bigg)+\NGen^2 \bigg(\frac{10549 \ala^2 \NR^2}{243000}+\frac{1519 \ala^2
   \NR}{4500} \notag \\
 & +\frac{567 \ala^2}{1000}+\frac{11}{900} \ala \alb \NR^2+\frac{7 \ala \alb
   \NR}{50}+\frac{27 \ala \alb}{100}+\frac{121}{675} \ala \alc \cR \NR^2+\frac{11}{25}
   \ala \alc \cR \NR \notag \\
 & +\frac{\alb^2 \NR^2}{360}+\frac{\alb^2
   \NR}{36}+\frac{\alb^2}{40}+\frac{44}{135} \alc^2 \cR \ItoR \NR\bigg)+\frac{21
   \ala^2}{1000}+\frac{9 \ala \alb}{100}-\frac{7 \ala \NR \YtD}{720} \notag \\
 & +\frac{17 \ala
   \NR \YtU}{720}+\frac{33 \ala \YtL}{80}-\frac{43 \alb^2}{40}+\frac{\alb \NR
   \YtD}{16}+\frac{17 \alb \NR \YtU}{80}+\frac{9 \alb \YtL}{16} \notag \\
 & +\frac{1}{6} \alc   \cR \NR \YtD+\frac{17}{30} \alc \cR \NR \YtU-\frac{\NR^2
   \YtD^2}{36}-\frac{11}{90} \NR^2 \YtD \YtU-\frac{17 \NR^2 \YtU^2}{180} \notag \\
 & -\frac{5 \NR
   \YtD \YtL}{18}-\frac{\NR \YtDs}{24}-\frac{31 \NR \YtL \YtU}{90}+\frac{11 \NR
   \YtUD}{60}-\frac{17 \NR \YtUs}{120}-\frac{\YtL^2}{4}-\frac{3 \YtLs}{8}\bigg] \notag \\
 &  + \invep\bigg[\NGen \bigg(-\frac{1697 \ala^2 \NR}{54000}-\frac{327 \ala^2}{2000}-\frac{\ala \alb
   \NR}{3600}-\frac{9 \ala \alb}{400}-\frac{137 \ala \alc \cR
   \NR}{2700} \notag \\
 & +\frac{\alb^2 \NR}{135}+\frac{9 \alb^2}{10}-\frac{1}{60} \alb \alc \cR
   \NR+\frac{1463 \alc^2 \cA \cR \NR}{1620}-\frac{11}{90} \alc^2 \cR^2
   \NR\bigg) \notag \\
 & +\NGen^2 \bigg(-\frac{16577 \ala^2 \NR^2}{1458000}-\frac{2387 \ala^2
   \NR}{27000}-\frac{297 \ala^2}{2000}-\frac{11 \alb^2 \NR^2}{2160}-\frac{11 \alb^2
   \NR}{216}-\frac{11 \alb^2}{240} \notag \\
 & -\frac{242}{405} \alc^2 \cR \ItoR \NR\bigg)+\frac{163
   \ala^2}{8000}+\frac{261 \ala \alb}{800}+\frac{9 \ala \lam}{50}-\frac{1267 \ala \NR
   \YtD}{7200} \notag \\
 & -\frac{2827 \ala \NR \YtU}{7200}-\frac{843 \ala \YtL}{800}+\frac{3401
   \alb^2}{960}+\frac{3 \alb \lam}{10}-\frac{437 \alb \NR \YtD}{480}-\frac{157 \alb
   \NR \YtU}{96} \notag \\
 & -\frac{543 \alb \YtL}{160}-\frac{17}{60} \alc \cR \NR
   \YtD-\frac{29}{60} \alc \cR \NR \YtU-\frac{3 \lam^2}{5}+\frac{17 \NR^2
   \YtD^2}{360}+\frac{59}{180} \NR^2 \YtD \YtU \notag \\
 & +\frac{101 \NR^2 \YtU^2}{360}+\frac{157
   \NR \YtD \YtL}{180}+\frac{61 \NR \YtDs}{240}+\frac{199 \NR \YtL
   \YtU}{180}+\frac{\NR \YtUD}{24}+\frac{113 \NR \YtUs}{240} \notag \\
 & +\frac{33 \YtL^2}{40}+\frac{87
   \YtLs}{80}\bigg]\bigg\}\,,
\end{align}

\begin{align}
Z_{\alpha_2} & = 1 + \alb\invep\bigg\{\NGen \bigg(\frac{\NR}{3}+\frac{1}{3}\bigg)-\frac{43}{6}\bigg\} \notag \\
 & + \alb\bigg\{\epp\bigg[\NGen^2 \bigg(\frac{\alb \NR^2}{9}+\frac{2 \alb \NR}{9}+\frac{\alb}{9}\bigg)+\NGen
   \bigg(-\frac{43 \alb \NR}{9}-\frac{43 \alb}{9}\bigg)+\frac{1849
     \alb}{36} \bigg] \notag \\
 &  + \invep\bigg[\NGen \bigg(\frac{\ala \NR}{120}+\frac{3 \ala}{40}+\frac{49 \alb \NR}{24}+\frac{49
   \alb}{24}+\frac{\alc \cR \NR}{2}\bigg) \notag \\
  & +\frac{3 \ala}{20}-\frac{259
   \alb}{12}-\frac{\NR \YtD}{4}-\frac{\NR \YtU}{4}-\frac{\YtL}{4} \bigg]\bigg\} \notag \\
  & + \alb\bigg\{\eppp\bigg[\NGen \bigg(\frac{1849 \alb^2 \NR}{36}+\frac{1849
   \alb^2}{36}\bigg)+\NGen^2 \bigg(-\frac{43}{18} \alb^2 \NR^2-\frac{43 \alb^2
   \NR}{9}-\frac{43 \alb^2}{18}\bigg) \notag \\
  & +\NGen^3 \bigg(\frac{\alb^2 \NR^3}{27}+\frac{\alb^2 \NR^2}{9}+\frac{\alb^2
   \NR}{9}+\frac{\alb^2}{27}\bigg)-\frac{79507 \alb^2}{216} \bigg] + \epp\bigg[ \NGen \bigg(\frac{\ala^2 \NR}{80}+\frac{13 \ala^2}{400} \notag \\
  & -\frac{7 \ala \alb   \NR}{360}-\frac{39 \ala \alb}{40}-\frac{22001 \alb^2 \NR}{432}-\frac{22001
   \alb^2}{432}-\frac{43}{6} \alb \alc \cR \NR-\frac{1}{6} \alb \NR^2
   \YtD \notag \\
  & -\frac{1}{6} \alb \NR^2 \YtU-\frac{\alb \NR \YtD}{6}-\frac{\alb \NR
   \YtL}{6}-\frac{\alb \NR \YtU}{6}-\frac{\alb \YtL}{6}-\frac{11}{18} \alc^2 \cA
   \cR \NR\bigg) \notag \\
  & +\NGen^2 \bigg(\frac{11 \ala^2 \NR^2}{16200}+\frac{7 \ala^2
   \NR}{900}+\frac{3 \ala^2}{200}+\frac{1}{180} \ala \alb \NR^2+\frac{\ala \alb
   \NR}{18}+\frac{\ala \alb}{20}+\frac{343 \alb^2 \NR^2}{216} \notag \\
  & +\frac{343 \alb^2
   \NR}{108}+\frac{343 \alb^2}{216}+\frac{1}{3} \alb \alc \cR \NR^2+\frac{1}{3} \alb
   \alc \cR \NR+\frac{4}{9} \alc^2 \cR \ItoR
   \NR\bigg)+\frac{\ala^2}{200} \notag \\
  & -\frac{43 \ala \alb}{20}+\frac{\ala \NR
   \YtD}{48}+\frac{17 \ala \NR \YtU}{240}+\frac{3 \ala \YtL}{16}+\frac{77959
   \alb^2}{216}+\frac{181 \alb \NR \YtD}{48} \notag \\
  & +\frac{181 \alb \NR \YtU}{48}+\frac{181
   \alb \YtL}{48}+\frac{1}{2} \alc \cR \NR \YtD+\frac{1}{2} \alc \cR \NR
   \YtU-\frac{\NR^2 \YtD^2}{12}-\frac{1}{6} \NR^2 \YtD \YtU \notag \\
  & -\frac{\NR^2
   \YtU^2}{12}-\frac{\NR \YtD \YtL}{6}-\frac{\NR \YtDs}{8}-\frac{\NR \YtL
   \YtU}{6}+\frac{\NR \YtUD}{4}-\frac{\NR \YtUs}{8}-\frac{\YtL^2}{12}-\frac{\YtLs}{8}\bigg]  \notag \\
  & + \invep\bigg[ \NGen \bigg(-\frac{287 \ala^2 \NR}{10800}-\frac{91 \ala^2}{1200}+\frac{13 \ala \alb
   \NR}{720}+\frac{13 \ala \alb}{80}-\frac{1}{180} \ala \alc \cR \NR \notag \\
  & +\frac{1603
   \alb^2 \NR}{81}+\frac{1603 \alb^2}{81}+\frac{13}{12} \alb \alc \cR
   \NR+\frac{133}{108} \alc^2 \cA \cR \NR-\frac{1}{6} \alc^2 \cR^2
   \NR\bigg) \notag \\
  & +\NGen^2 \bigg(-\frac{121 \ala^2 \NR^2}{97200}-\frac{77 \ala^2 \NR}{5400}-\frac{11
   \ala^2}{400}-\frac{415 \alb^2 \NR^2}{1296}-\frac{415 \alb^2 \NR}{648}-\frac{415
   \alb^2}{1296} \notag \\
  & -\frac{22}{27} \alc^2 \cR \ItoR \NR\bigg)+\frac{163 \ala^2}{4800}+\frac{187
   \ala \alb}{160}+\frac{\ala \lam}{10}-\frac{533 \ala \NR \YtD}{1440}-\frac{593
   \ala \NR \YtU}{1440} \notag \\
  & -\frac{17 \ala \YtL}{32}-\frac{667111 \alb^2}{5184}+\frac{\alb
   \lam}{2}-\frac{81 \alb \NR \YtD}{32}-\frac{81 \alb \NR \YtU}{32}-\frac{81 \alb
   \YtL}{32} \notag \\
  & -\frac{7}{12} \alc \cR \NR \YtD-\frac{7}{12} \alc \cR \NR
   \YtU-\lam^2+\frac{5 \NR^2 \YtD^2}{24}+\frac{5}{12} \NR^2 \YtD \YtU+\frac{5
   \NR^2 \YtU^2}{24} \notag \\
  & +\frac{5 \NR \YtD \YtL}{12}+\frac{19 \NR \YtDs}{48}+\frac{5
   \NR \YtL \YtU}{12}+\frac{3 \NR \YtUD}{8}+\frac{19 \NR \YtUs}{48} \notag \\
  & +\frac{5
   \YtL^2}{24}+\frac{19 \YtLs}{48}\bigg]\bigg\}\,,
\end{align}

\begin{align}
Z_{\alpha_s} & = 1 + \alc\invep\bigg\{\frac{8 \ItoR \NGen}{3}-\frac{11 \cA}{3}\bigg\} \notag \\
 & + \alc\bigg\{\epp\bigg[\frac{121 \alc \cA^2}{9}-\frac{176}{9} \alc \cA \ItoR \NGen+\frac{64}{9} \alc \ItoR^2
   \NGen^2 \bigg] \notag \\
 & + \invep\bigg[\NGen \bigg(\frac{11 \ala \ItoR}{30}+\frac{3 \alb \ItoR}{2}+\frac{20 \alc \cA
   \ItoR}{3}+4 \alc \cR \ItoR\bigg) \notag \\
   & -\frac{17 \alc \cA^2}{3}-2 \ItoR \YtD-2\ItoR \YtU \bigg]\bigg\} \notag \\
  & + \alc\bigg\{\eppp\bigg[ -\frac{1331}{27} \alc^2 \cA^3+\frac{968}{9} \alc^2 \cA^2 \ItoR \NGen-\frac{704}{9} \alc^2
   \cA \ItoR^2 \NGen^2+\frac{512}{27} \alc^2 \ItoR^3 \NGen^3\bigg] \notag \\
  &  + \epp\bigg[\NGen \bigg(\frac{11 \ala^2 \ItoR}{900}-\frac{121}{45} \ala \alc \cA \ItoR-\frac{43
   \alb^2 \ItoR}{12}-11 \alb \alc \cA \ItoR-\frac{2492}{27} \alc^2 \cA^2
   \ItoR \notag \\
  & -\frac{308}{9} \alc^2 \cA \cR \ItoR-\frac{32}{3} \alc \ItoR^2
   \YtD-\frac{32}{3} \alc \ItoR^2 \YtU\bigg)+\NGen^2 \bigg(\frac{121 \ala^2 \ItoR
   \NR}{4050}+\frac{11 \ala^2 \ItoR}{150} \notag \\
  & +\frac{88}{45} \ala \alc \ItoR^2+\frac{1}{6}
   \alb^2 \ItoR \NR+\frac{\alb^2 \ItoR}{6}+8 \alb \alc \ItoR^2+\frac{1120}{27}
   \alc^2 \cA \ItoR^2+\frac{224}{9} \alc^2 \cR \ItoR^2\bigg) \notag \\
  & +\frac{\ala \ItoR
   \YtD}{6}+\frac{17 \ala \ItoR \YtU}{30}+\frac{3 \alb \ItoR \YtD}{2}+\frac{3 \alb
   \ItoR \YtU}{2}+\frac{1309 \alc^2 \cA^3}{27}+\frac{44}{3} \alc \cA \ItoR
   \YtD \notag \\
  & +\frac{44}{3} \alc \cA \ItoR \YtU+4 \alc \cR \ItoR \YtD+4 \alc
   \cR \ItoR \YtU-\frac{2}{3} \ItoR \NR \YtD^2-\frac{4}{3} \ItoR \NR \YtD
   \YtU \notag \\
  & -\frac{2}{3} \ItoR \NR \YtU^2-\frac{2 \ItoR \YtD \YtL}{3}-\ItoR
   \YtDs-\frac{2 \ItoR \YtL \YtU}{3}+2 \ItoR \YtUD-\ItoR \YtUs \bigg] \notag \\
  &  + \invep\bigg[\NGen \bigg(-\frac{13 \ala^2 \ItoR}{180}-\frac{\ala \alb \ItoR}{60}+\frac{22}{45} \ala
   \alc \cA \ItoR-\frac{11}{45} \ala \alc \cR \ItoR+\frac{241 \alb^2\ItoR}{36} \notag \\
  & +2 \alb \alc \cA \ItoR-\alb \alc \cR \ItoR+\frac{2830}{81}
   \alc^2 \cA^2 \ItoR+\frac{410}{27} \alc^2 \cA \cR \ItoR-\frac{4}{3} \alc^2
   \cR^2 \ItoR\bigg) \notag \\
   &+\NGen^2 \bigg(-\frac{1331 \ala^2 \ItoR \NR}{24300} -\frac{121 \ala^2
   \ItoR}{900}-\frac{11}{36} \alb^2 \ItoR \NR-\frac{11 \alb^2 \ItoR}{36}-\frac{632}{81}
   \alc^2 \cA \ItoR^2 \notag \\
  & -\frac{176}{27} \alc^2 \cR \ItoR^2\bigg)-\frac{89 \ala \ItoR
   \YtD}{60}-\frac{101 \ala \ItoR \YtU}{60}-\frac{31 \alb \ItoR \YtD}{4}-\frac{31
   \alb \ItoR \YtU}{4} \notag \\
  & -\frac{2857}{162} \alc^2 \cA^3-8 \alc \cA \ItoR \YtD-8
   \alc \cA \ItoR \YtU-2 \alc \cR \ItoR \YtD-2 \alc \cR \ItoR
   \YtU \notag \\
  & +\frac{7}{3} \ItoR \NR \YtD^2+\frac{14}{3} \ItoR \NR \YtD \YtU+\frac{7}{3}
   \ItoR \NR \YtU^2+\frac{7 \ItoR \YtD \YtL}{3}+3 \ItoR \YtDs+\frac{7 \ItoR
   \YtL \YtU}{3} \notag \\
   &-2 \ItoR \YtUD+3 \ItoR \YtUs \bigg]\bigg\}\,,
\end{align}

\begin{align}
Z_{\xi_B} & = 1 + \ala\invep\bigg\{\NGen \bigg(-\frac{11 \NR}{45}-\frac{3}{5}\bigg)-\frac{1}{10}\bigg\} \notag \\
 & + \ala\bigg\{ \invep\bigg[ \NGen \bigg(-\frac{137 \ala \NR}{1800}-\frac{81 \ala}{200}-\frac{\alb \NR}{40}-\frac{9
   \alb}{40}-\frac{11 \alc \cR \NR}{30}\bigg) \notag \\
  & -\frac{9 \ala}{100}-\frac{9
   \alb}{20}+\frac{\NR \YtD}{12}+\frac{17 \NR \YtU}{60}+\frac{3 \YtL}{4}\bigg]\bigg\} \notag \\
  & + \ala\bigg\{ \epp\bigg[ \NGen \bigg(-\frac{533 \ala^2 \NR}{54000}-\frac{63 \ala^2}{2000}+\frac{7 \alb^2
   \NR}{720}+\frac{39 \alb^2}{80}+\frac{121}{270} \alc^2 \cA \cR \NR\bigg) \notag \\
  & -\NGen^2
   \bigg(\frac{1507 \ala^2 \NR^2}{243000}+\frac{217 \ala^2 \NR}{4500}+\frac{81
   \ala^2}{1000}+\frac{\alb^2 \NR^2}{360}+\frac{\alb^2
   \NR}{36}+\frac{\alb^2}{40}+\frac{44}{135} \alc^2 \cR \ItoR \NR\bigg) \notag \\
  & -\frac{3
   \ala^2}{1000}-\frac{\ala \NR \YtD}{144}-\frac{289 \ala \NR \YtU}{3600}-\frac{9
   \ala \YtL}{16}+\frac{43 \alb^2}{40}-\frac{\alb \NR \YtD}{16}-\frac{17 \alb \NR
   \YtU}{80} \notag \\
  & -\frac{9 \alb \YtL}{16}-\frac{1}{6} \alc \cR \NR \YtD-\frac{17}{30}
   \alc \cR \NR \YtU+\frac{\NR^2 \YtD^2}{36}+\frac{11}{90} \NR^2 \YtD
   \YtU+\frac{17 \NR^2 \YtU^2}{180} \notag \\
  & +\frac{5 \NR \YtD \YtL}{18}+\frac{\NR
   \YtDs}{24}+\frac{31 \NR \YtL \YtU}{90}-\frac{11 \NR \YtUD}{60}+\frac{17 \NR
   \YtUs}{120}+\frac{\YtL^2}{4}+\frac{3 \YtLs}{8}\bigg] \notag \\
  &  + \invep\bigg[\NGen \bigg(\frac{1697 \ala^2 \NR}{54000}+\frac{327 \ala^2}{2000}+\frac{\ala \alb
   \NR}{3600}+\frac{9 \ala \alb}{400}+\frac{137 \ala \alc \cR
   \NR}{2700}-\frac{\alb^2 \NR}{135} \notag \\
  & -\frac{9 \alb^2}{10}+\frac{1}{60} \alb \alc \cR
   \NR-\frac{1463 \alc^2 \cA \cR \NR}{1620}+\frac{11}{90} \alc^2 \cR^2
   \NR\bigg)+\NGen^2 \bigg(\frac{16577 \ala^2 \NR^2}{1458000} \notag \\
  & +\frac{2387 \ala^2
   \NR}{27000}+\frac{297 \ala^2}{2000}+\frac{11 \alb^2 \NR^2}{2160}+\frac{11 \alb^2
   \NR}{216}+\frac{11 \alb^2}{240}+\frac{242}{405} \alc^2 \cR \ItoR \NR\bigg) \notag \\
  & -\frac{163
   \ala^2}{8000}-\frac{261 \ala \alb}{800}-\frac{9 \ala \lam}{50}+\frac{1267 \ala \NR
   \YtD}{7200}+\frac{2827 \ala \NR \YtU}{7200}+\frac{843 \ala \YtL}{800} \notag \\
  & -\frac{3401
   \alb^2}{960}-\frac{3 \alb \lam}{10}+\frac{437 \alb \NR \YtD}{480}+\frac{157 \alb
   \NR \YtU}{96}+\frac{543 \alb \YtL}{160}+\frac{17}{60} \alc \cR \NR
   \YtD \notag \\
  & +\frac{29}{60} \alc \cR \NR \YtU+\frac{3 \lam^2}{5}-\frac{17 \NR^2
   \YtD^2}{360}-\frac{59}{180} \NR^2 \YtD \YtU-\frac{101 \NR^2 \YtU^2}{360}-\frac{33 \YtL^2}{40} \notag \\
  & -\frac{157
   \NR \YtD \YtL}{180}-\frac{61 \NR \YtDs}{240}-\frac{199 \NR \YtL
   \YtU}{180}-\frac{\NR \YtUD}{24} \notag \\
  & -\frac{113 \NR \YtUs}{240}-\frac{87
   \YtLs}{80} \bigg]\bigg\}\,,
\end{align}

\begin{align}
Z_{\xi_W} & = 1 + \alb\invep\bigg\{-\xiW+\NGen \bigg(-\frac{\NR}{3}-\frac{1}{3}\bigg)+\frac{25}{6}\bigg\} \notag \\
 & + \alb\bigg\{\epp\bigg[ \alb \xiW^2+\NGen \bigg(\frac{\alb \xiW \NR}{3}+\frac{\alb
   \xiW}{3}+\frac{\alb \NR}{2}+\frac{\alb}{2}\bigg)-\frac{8 \alb \xiW}{3}-\frac{25
   \alb}{4}\bigg]  \notag \\
  & + \invep\bigg[\NGen \bigg(-\frac{\ala \NR}{120}-\frac{3 \ala}{40}-\frac{13 \alb \NR}{8}-\frac{13
   \alb}{8}-\frac{\alc \cR \NR}{2}\bigg) \notag \\
  & -\frac{3 \ala}{20}-\frac{\alb
   \xiW^2}{2}-\frac{11 \alb \xiW}{4}+\frac{113 \alb}{8}+\frac{\NR
   \YtD}{4}+\frac{\NR \YtU}{4}+\frac{\YtL}{4} \bigg]\bigg\} \notag \\
  & + \alb\bigg\{\eppp\bigg[-\alb^2 \xiW^3+\NGen \bigg(-\frac{1}{3} \alb^2 \xiW^2 \NR-\frac{1}{3} \alb^2
   \xiW^2-\frac{5}{6} \alb^2 \xiW \NR-\frac{5 \alb^2 \xiW}{6} \notag \\
  &                  -\frac{43
   \alb^2 \NR}{18}-\frac{43 \alb^2}{18}\bigg)+\frac{7 \alb^2 \xiW^2}{6}+\frac{89 \alb^2
   \xiW}{12}+\NGen^2 \bigg(\frac{\alb^2 \NR^2}{18}+\frac{\alb^2
   \NR}{9}+\frac{\alb^2}{18}\bigg) \notag \\
  &                  +\frac{1525 \alb^2}{72} \bigg] + \epp\bigg[ \NGen \bigg(-\frac{\ala^2 \NR}{80}-\frac{13 \ala^2}{400}+\frac{1}{120} \ala \alb
   \xiW \NR+\frac{3 \ala \alb \xiW}{40} \notag \\
  &                  +\frac{\ala \alb
   \NR}{120}+\frac{3 \ala \alb}{40}+\frac{1}{6} \alb^2 \xiW^2 \NR+\frac{\alb^2
   \xiW^2}{6}+\frac{47}{24} \alb^2 \xiW \NR+\frac{47 \alb^2
   \xiW}{24} \notag \\
  &                 +\frac{4273 \alb^2 \NR}{432}+\frac{4273 \alb^2}{432}+\frac{1}{2} \alb \alc
   \cR \xiW \NR+\frac{1}{2} \alb \alc \cR \NR+\frac{11}{18} \alc^2 \cA
   \cR \NR\bigg) \notag \\
   &+\NGen^2 \bigg(-\frac{11 \ala^2 \NR^2}{16200}-\frac{7 \ala^2
   \NR}{900}-\frac{3 \ala^2}{200}-\frac{59 \alb^2 \NR^2}{216}-\frac{59 \alb^2
   \NR}{108}-\frac{59 \alb^2}{216} \notag \\
  &                 -\frac{4}{9} \alc^2 \cR \ItoR
   \NR\bigg)-\frac{\ala^2}{200}+\frac{3 \ala \alb \xiW}{20}+\frac{3 \ala
   \alb}{20}-\frac{\ala \NR \YtD}{48}-\frac{17 \ala \NR \YtU}{240} \notag \\
  &                 -\frac{3 \ala\YtL}{16}+\frac{7 \alb^2 \xiW^3}{6}+\frac{53 \alb^2 \xiW^2}{12}-\frac{271 \alb^2
   \xiW}{24}-\frac{29629 \alb^2}{432}-\frac{1}{4} \alb \xiW \NR \YtD \notag \\
  &                 -\frac{1}{4} \alb \xiW \NR \YtU-\frac{\alb \xiW \YtL}{4}-\frac{7 \alb \NR
   \YtD}{16}-\frac{7 \alb \NR \YtU}{16}-\frac{7 \alb \YtL}{16}-\frac{1}{2} \alc
   \cR \NR \YtD \notag \\
  &                 -\frac{1}{2} \alc \cR \NR \YtU+\frac{\NR^2
   \YtD^2}{12}+\frac{1}{6} \NR^2 \YtD \YtU+\frac{\NR^2 \YtU^2}{12}+\frac{\NR \YtD
   \YtL}{6}+\frac{\NR \YtDs}{8} \notag \\
  &                 +\frac{\NR \YtL \YtU}{6}-\frac{\NR
   \YtUD}{4}+\frac{\NR \YtUs}{8}+\frac{\YtL^2}{12}+\frac{\YtLs}{8}\bigg]+ \invep\bigg[ \NGen \bigg(\frac{287 \ala^2 \NR}{10800}+\frac{91 \ala^2}{1200} \notag \\
  &                 -\frac{1}{15} \ala \alb
   \NR \zetathree+\frac{2 \ala \alb \NR}{45}-\frac{3 \ala \alb \zetathree}{5}+\frac{2
   \ala \alb}{5}+\frac{1}{180} \ala \alc \cR \NR \notag \\
  &                 +\frac{2}{3} \alb^2 \xiW
   \NR+\frac{2 \alb^2 \xiW}{3}+3 \alb^2 \NR \zetathree-\frac{7025 \alb^2
   \NR}{432}+3 \alb^2 \zetathree-\frac{7025 \alb^2}{432} \notag \\
  &                -4 \alb \alc \cR \NR
   \zetathree+\frac{8}{3} \alb \alc \cR \NR-\frac{133}{108} \alc^2 \cA \cR
   \NR+\frac{1}{6} \alc^2 \cR^2 \NR\bigg) \notag \\
  &                +\NGen^2 \bigg(\frac{121 \ala^2
   \NR^2}{97200}+\frac{77 \ala^2 \NR}{5400}+\frac{11 \ala^2}{400}+\frac{185 \alb^2
   \NR^2}{432}+\frac{185 \alb^2 \NR}{216} +\frac{185 \alb^2}{432} \notag \\
  & +\frac{22}{27} \alc^2 \cR
   \ItoR \NR\bigg)-\frac{163 \ala^2}{4800}-\frac{3 \ala \alb \zetathree}{10}-\frac{11 \ala
   \alb}{32}-\frac{ \ala \lam}{10}+\frac{533 \ala \NR \YtD}{1440} \notag \\
 & +\frac{593 \ala
   \NR \YtU}{1440}+\frac{17 \ala \YtL}{32}-\frac{7}{12} \alb^2 \xiW^3-\frac{1}{2}
   \alb^2 \xiW^2 \zetathree-\frac{11 \alb^2 \xiW^2}{4}-2 \alb^2 \xiW
   \zetathree \notag \\
  & -\frac{105 \alb^2 \xiW}{8}+\frac{\alb^2 \zetathree}{2}+\frac{143537 \alb^2}{1728}-
  \frac{\alb \lam}{2} +\frac{79 \alb \NR \YtD}{96}+\frac{79 \alb \NR \YtU}{96} \notag \\
  & +\frac{79
   \alb \YtL}{96}+\frac{7}{12} \alc \cR \NR \YtD+\frac{7}{12} \alc \cR \NR
   \YtU+ \lam^2
   -\frac{5 \NR^2 \YtD^2}{24}-\frac{5}{12} \NR^2 \YtD \YtU \notag \\
  & -\frac{5
   \NR^2 \YtU^2}{24}-\frac{5 \NR \YtD \YtL}{12}-\frac{19 \NR \YtDs}{48}-\frac{5
   \NR \YtL \YtU}{12}
   -\frac{3 \NR \YtUD}{8}-\frac{19 \NR \YtUs}{48} \notag \\
  & -\frac{5
   \YtL^2}{24}-\frac{19 \YtLs}{48}\bigg]\bigg\}\,,
\end{align}

\begin{align}
Z_{\xi_G} & = 1 + \alc\invep\bigg\{-\frac{\cA \xiG}{2}+\frac{13 \cA}{6}-\frac{8 \ItoR \NGen}{3}\bigg\} \notag \\
 & + \alc\bigg\{\epp\bigg[ \frac{1}{4} \alc \cA^2 \xiG^2-\frac{17}{24} \alc \cA^2 \xiG-\frac{13 \alc
   \cA^2}{8} \notag \\
  & +\NGen \left(\frac{4}{3} \alc \cA \xiG \ItoR+2 \alc \cA
   \ItoR\right)\bigg] \notag \\
  &  + \invep\bigg[\NGen \bigg(-\frac{11 \ala \ItoR}{30}-\frac{3 \alb \ItoR}{2}-5 \alc \cA \ItoR-4
   \alc \cR \ItoR\bigg) \notag \\
 & -\frac{1}{8} \alc \cA^2 \xiG^2-\frac{11}{16} \alc
   \cA^2 \xiG+\frac{59 \alc \cA^2}{16}+2 \ItoR \YtD+2 \ItoR \YtU \bigg]\bigg\} \notag \\
  & + \alc\bigg\{\eppp\bigg[-\frac{1}{8} \alc^2 \cA^3 \xiG^3+\frac{1}{6} \alc^2 \cA^3 \xiG^2+\frac{47}{48}
   \alc^2 \cA^3 \xiG+\frac{403 \alc^2 \cA^3}{144} \notag \\
  & +\NGen \bigg(-\frac{2}{3} \alc^2
   \cA^2 \xiG^2 \ItoR-\frac{5}{3} \alc^2 \cA^2 \xiG \ItoR-\frac{44}{9}
   \alc^2 \cA^2 \ItoR\bigg)+\frac{16}{9} \alc^2 \cA \ItoR^2 \NGen^2 \bigg] \notag \\
  & + \epp\bigg[ \NGen \bigg(-\frac{11 \ala^2 \ItoR}{900} +\frac{11}{60} \ala \alc \cA \xiG
   \ItoR+\frac{11}{60} \ala \alc \cA \ItoR+\frac{43 \alb^2 \ItoR}{12} \notag \\
  & +\frac{3}{4}
   \alb \alc \cA \xiG \ItoR+\frac{3}{4} \alb \alc \cA \ItoR +\frac{1}{3}
   \alc^2 \cA^2 \xiG^2 \ItoR+\frac{19}{6} \alc^2 \cA^2 \xiG
   \ItoR \notag \\
  & +\frac{481}{27} \alc^2 \cA^2 \ItoR+2 \alc^2 \cA \cR \xiG
   \ItoR+\frac{62}{9} \alc^2 \cA \cR \ItoR\bigg)+\NGen^2 \bigg(-\frac{121 \ala^2
   \ItoR \NR}{4050} \notag \\
  & -\frac{11 \ala^2 \ItoR}{150}-\frac{1}{6} \alb^2 \ItoR
   \NR-\frac{\alb^2 \ItoR}{6}-\frac{200}{27} \alc^2 \cA \ItoR^2-\frac{32}{9} \alc^2
   \cR \ItoR^2\bigg)-\frac{\ala \ItoR \YtD}{6} \notag \\
  & -\frac{17 \ala \ItoR
   \YtU}{30}-\frac{3 \alb \ItoR \YtD}{2}-\frac{3 \alb \ItoR \YtU}{2}+\frac{7}{48}
   \alc^2 \cA^3 \xiG^3+\frac{13}{24} \alc^2 \cA^3 \xiG^2 \notag \\
  & -\frac{143}{96}
   \alc^2 \cA^3 \xiG-\frac{7957 \alc^2 \cA^3}{864}-\alc \cA \xiG
   \ItoR \YtD-\alc \cA \xiG \ItoR \YtU-\alc \cA \ItoR
   \YtD \notag \\
  & -\alc \cA \ItoR \YtU-4 \alc \cR \ItoR \YtD-4 \alc \cR
   \ItoR \YtU+\frac{2}{3} \ItoR \NR \YtD^2+\frac{4}{3} \ItoR \NR \YtD
   \YtU \notag \\
  & +\frac{2}{3} \ItoR \NR \YtU^2+\frac{2 \ItoR \YtD \YtL}{3}+\ItoR
   \YtDs+\frac{2 \ItoR \YtL \YtU}{3}-2 \ItoR \YtUD+\ItoR \YtUs\bigg] \notag \\
  &  + \invep\bigg[ \NGen \bigg(\frac{13 \ala^2 \ItoR}{180}+\frac{\ala \alb \ItoR}{60}-\frac{22}{15} \ala
   \alc \cA \ItoR \zetathree+\frac{319}{360} \ala \alc \cA \ItoR\notag\\
  & +\frac{11}{45} \ala
   \alc \cR \ItoR-\frac{241 \alb^2 \ItoR}{36}-6 \alb \alc \cA \ItoR
   \zetathree+\frac{29}{8} \alb \alc \cA \ItoR+\alb \alc \cR \ItoR\notag\\
  & +\frac{4}{3}
   \alc^2 \cA^2 \xiG \ItoR+12 \alc^2 \cA^2 \ItoR \zetathree-\frac{911}{27}
   \alc^2 \cA^2 \ItoR-16 \alc^2 \cA \cR \ItoR \zetathree\notag\\
  & -\frac{5}{27} \alc^2
   \cA \cR \ItoR+\frac{4}{3} \alc^2 \cR^2 \ItoR\bigg)+\NGen^2 \bigg(\frac{1331
   \ala^2 \ItoR \NR}{24300}+\frac{121 \ala^2 \ItoR}{900}+\frac{11}{36} \alb^2 \ItoR
   \NR\notag\\
  &  +\frac{11 \alb^2 \ItoR}{36}+\frac{304}{27} \alc^2 \cA \ItoR^2+\frac{176}{27}+\alc^2
   \cR \ItoR^2\bigg)+\frac{89 \ala \ItoR \YtD}{60}+\frac{101 \ala \ItoR
   \YtU}{60}\notag\\
  & +\frac{31 \alb \ItoR \YtD}{4}+\frac{31 \alb \ItoR \YtU}{4}-\frac{7}{96}
   \alc^2 \cA^3 \xiG^3-\frac{1}{16} \alc^2 \cA^3 \xiG^2 \zetathree-\frac{11}{32}
   \alc^2 \cA^3 \xiG^2\notag\\
  & -\frac{1}{4} \alc^2 \cA^3 \xiG \zetathree-\frac{167}{96}
   \alc^2 \cA^3 \xiG-\frac{3}{16} \alc^2 \cA^3 \zetathree+\frac{9965 \alc^2
   \cA^3}{864}+\frac{25}{6} \alc \cA \ItoR \YtD\notag\\
  & +\frac{25}{6} \alc \cA \ItoR
   \YtU+2 \alc \cR \ItoR \YtD+2 \alc \cR \ItoR \YtU-\frac{7}{3} \ItoR
   \NR \YtD^2-\frac{14}{3} \ItoR \NR \YtD \YtU\notag\\
  & -\frac{7}{3} \ItoR \NR
   \YtU^2-\frac{7 \ItoR \YtD \YtL}{3}-3 \ItoR \YtDs-\frac{7 \ItoR \YtL
   \YtU}{3} \notag \\
  & +2 \ItoR \YtUD-3 \ItoR \YtUs\bigg]\bigg\}\,.
\end{align}
}



\begin{thebibliography}{100}
\bibitem{:2012gk}
    [ATLAS Collaboration],
  ``Observation of a new particle in the search for the Standard Model Higgs
  boson with the ATLAS detector at the LHC,''
  Phys.\ Lett.\  B {\bf 716} (2012) 1
  [arXiv:1207.7214 [hep-ex]].

\bibitem{:2012gu}
    [CMS Collaboration],
  ``Observation of a new boson at a mass of 125 GeV with the CMS
  experiment at the LHC,''
  Phys.\ Lett.\  B {\bf 716} (2012) 30
  [arXiv:1207.7235 [hep-ex]].

\bibitem{Krasnikov:1978pu}
  N.~V.~Krasnikov,
  ``Restriction of the Fermion Mass in Gauge Theories of Weak and Electromagnetic Interactions,''
  Yad.\ Fiz.\  {\bf 28} (1978) 549.

\bibitem{Hung:1979dn}
  P.~Q.~Hung,
  ``Vacuum Instability and New Constraints on Fermion Masses,''
  Phys.\ Rev.\ Lett.\  {\bf 42} (1979) 873.

\bibitem{Politzer:1978ic}
  H.~D.~Politzer and S.~Wolfram,
  ``Bounds on Particle Masses in the Weinberg-Salam Model,''
  Phys.\ Lett.\ B {\bf 82} (1979) 242
   [Erratum-ibid.\  {\bf 83B} (1979) 421].

\bibitem{Bezrukov:2012sa}
  F.~Bezrukov, M.~Y.~Kalmykov, B.~A.~Kniehl and M.~Shaposhnikov,
  ``Higgs Boson Mass and New Physics,''
  arXiv:1205.2893 [hep-ph].

\bibitem{Degrassi:2012ry}
  G.~Degrassi, S.~Di Vita, J.~Elias-Miro, J.~R.~Espinosa, G.~F.~Giudice,
G.~Isidori and A.~Strumia,
  ``Higgs mass and vacuum stability in the Standard Model at NNLO,''
  JHEP {\bf 1208} (2012) 098
  [arXiv:1205.6497 [hep-ph]].

\bibitem{Alekhin:2012py}
  S.~Alekhin, A.~Djouadi and S.~Moch,
  ``The top quark and Higgs boson masses and the stability of the
   electroweak vacuum,''
  Phys.\ Lett.\  B {\bf 716} (2012) 214
  [arXiv:1207.0980 [hep-ph]].

\bibitem{Collins:1984xc}
  J.~C.~Collins,
  ``Renormalization. An introduction to renormalization, the renormalization
  group, and the operator product expansion,''
{\it  Cambridge, Uk: Univ. Pr. (1984) 380p}

\bibitem{Vladimirov:1979zm}
  A.~A.~Vladimirov,
  ``Method for computing renormalization group functions in dimensional
  renormalization scheme,''
  Theor.\ Math.\ Phys.\  {\bf 43} (1980) 417
  [Teor.\ Mat.\ Fiz.\  {\bf 43} (1980) 210].

\bibitem{Gross:1973id}
  D.~J.~Gross and F.~Wilczek,
  ``Ultraviolet behavior of nonabelian gauge theories,''
  Phys.\ Rev.\ Lett.\  {\bf 30} (1973) 1343.

\bibitem{Politzer:1973fx}
  H.~D.~Politzer,
  ``Reliable perturbative results for strong interactions?,''
  Phys.\ Rev.\ Lett.\  {\bf 30} (1973) 1346.

\bibitem{Jones:1974mm}
  D.~R.~T.~Jones,
  ``Two loop diagrams in Yang-Mills theory,''
  Nucl.\ Phys.\  B {\bf 75} (1974) 531.

\bibitem{Tarasov:1976ef}
  O.~V.~Tarasov and A.~A.~Vladimirov,
  ``Two loop renormalization of the Yang-Mills theory in an arbitrary
  gauge,''
  Sov.\ J.\ Nucl.\ Phys.\  {\bf 25} (1977) 585
  [Yad.\ Fiz.\  {\bf 25} (1977) 1104].

\bibitem{Caswell:1974gg}
  W.~E.~Caswell,
  ``Asymptotic behavior of nonabelian gauge theories to two loop order,''
  Phys.\ Rev.\ Lett.\  {\bf 33} (1974) 244.

\bibitem{Egorian:1978zx}
  E.~Egorian and O.~V.~Tarasov,
  ``Two loop renormalization of the QCD in an arbitrary gauge,''
  Teor.\ Mat.\ Fiz.\  {\bf 41} (1979) 26
  [Theor.\ Math.\ Phys.\  {\bf 41} (1979) 863].

\bibitem{Jones:1981we}
  D.~R.~T.~Jones,
  ``The two loop beta function for a G(1) x G(2) gauge theory,''
  Phys.\ Rev.\  D {\bf 25} (1982) 581.

\bibitem{Fischler:1981is}
  M.~S.~Fischler and C.~T.~Hill,
  ``Effects of large mass fermions on $M_X$ and $\sin^2\theta_W$,''
  Nucl.\ Phys.\  B {\bf 193} (1981) 53.

\bibitem{Machacek:1983tz}
  M.~E.~Machacek and M.~T.~Vaughn,
  ``Two loop renormalization group equations in a general quantum field
  theory. 1. Wave function renormalization,''
  Nucl.\ Phys.\  B {\bf 222} (1983) 83.

\bibitem{Jack:1984vj}
  I.~Jack and H.~Osborn,
  ``General background field calculations with fermion fields,''
  Nucl.\ Phys.\  B {\bf 249} (1985) 472.

\bibitem{Gorishnii:1987ik}
  S.~G.~Gorishnii, A.~L.~Kataev and S.~A.~Larin,
  ``Two Loop Renormalization Group Calculations In Theories With Scalar Quarks,''
  Theor.\ Math.\ Phys.\  {\bf 70} (1987) 262
   [Teor.\ Mat.\ Fiz.\  {\bf 70} (1987) 372].

\bibitem{Arason:1991ic}
  H.~Arason, D.~J.~Castano, B.~Keszthelyi, S.~Mikaelian, E.~J.~Piard,
P.~Ramond and B.~D.~Wright,
  ``Renormalization group study of the standard model and its extensions. 1.
  The Standard model,''
  Phys.\ Rev.\  D {\bf 46} (1992) 3945.

\bibitem{Curtright:1979mg}
  T.~Curtright,
  ``Three loop charge renormalization effects due to quartic scalar
  selfinteractions,''
  Phys.\ Rev.\  D {\bf 21} (1980) 1543.

\bibitem{Jones:1980fx}
  D.~R.~T.~Jones,
  ``Comment on the charge renormalization effects of quartic scalar
  selfinteractions,''
  Phys.\ Rev.\  D {\bf 22} (1980) 3140.

\bibitem{Tarasov:1980au}
  O.~V.~Tarasov, A.~A.~Vladimirov and A.~Y.~Zharkov,
  ``The Gell-Mann-Low function of QCD in the three loop approximation,''
  Phys.\ Lett.\  B {\bf 93} (1980) 429.

\bibitem{Larin:1993tp}
  S.~A.~Larin and J.~A.~M.~Vermaseren,
  ``The three loop QCD Beta function and anomalous dimensions,''
  Phys.\ Lett.\  B {\bf 303} (1993) 334
  [arXiv:hep-ph/9302208].

\bibitem{Steinhauser:1998cm}
  M.~Steinhauser,
  ``Higgs decay into gluons up to $\mathcal{O}(\alpha_s^3 G_F m_t^2)$,''
  Phys.\ Rev.\  D {\bf 59} (1999) 054005
  [arXiv:hep-ph/9809507].

\bibitem{Pickering:2001aq}
  A.~G.~M.~Pickering, J.~A.~Gracey and D.~R.~T.~Jones,
  ``Three loop gauge beta function for the most general single gauge
  coupling theory,''
  Phys.\ Lett.\  B {\bf 510} (2001) 347
  [Phys.\ Lett.\  B {\bf 512} (2001) 230]
  [Erratum-ibid.\  B {\bf 535} (2002) 377]
  [arXiv:hep-ph/0104247].

\bibitem{Velizhanin:2008rw}
  V.~N.~Velizhanin,
  ``Three-loop renormalization of the N=1, N=2, N=4 supersymmetric
  Yang-Mills theories,''
  Nucl.\ Phys.\  B {\bf 818} (2009) 95
  [arXiv:0809.2509 [hep-th]].

\bibitem{Velizhanin:2012nm}
  V.~N.~Velizhanin,
  ``Three loop anomalous dimension of the non-singlet transversity
  operator in QCD,''
  Nucl.\ Phys.\  B {\bf 864} (2012) 113
  [arXiv:1203.1022 [hep-ph]].

\bibitem{Bagaev:2012bw}
  A.~A.~Bagaev, A.~V.~Bednyakov, A.~F.~Pikelner and V.~N.~Velizhanin,
  ``The 16th moment of the three loop anomalous dimension of the non-singlet
  transversity operator in QCD,''
  Phys.\ Lett.\  B {\bf 714} (2012) 76
  [arXiv:1206.2890 [hep-ph]].

\bibitem{Bednyakov:2007vm}
  A.~V.~Bednyakov,
  ``Running mass of the b-quark in QCD and SUSY QCD,''
  Int.\ J.\ Mod.\ Phys.\  A {\bf 22} (2007) 5245
  [arXiv:0707.0650 [hep-ph]].

\bibitem{Bednyakov:2009wt}
  A.~V.~Bednyakov,
  ``On the two-loop decoupling corrections to tau-lepton and b-quark running
  masses in the MSSM,''
  Int.\ J.\ Mod.\ Phys.\  A {\bf 25} (2010) 2437
  [arXiv:0912.4652 [hep-ph]].

\bibitem{Bednyakov:2010ni}
  A.~V.~Bednyakov,
  ``Some two-loop threshold corrections and three-loop renormalization group
  analysis of the MSSM,''
  arXiv:1009.5455 [hep-ph].

\bibitem{Machacek:1983fi}
  M.~E.~Machacek and M.~T.~Vaughn,
  ``Two loop renormalization group equations in a general quantum field
  theory. 2. Yukawa couplings,''
  Nucl.\ Phys.\  B {\bf 236} (1984) 221.

\bibitem{Machacek:1984zw}
  M.~E.~Machacek and M.~T.~Vaughn,
  ``Two loop renormalization group equations in a general quantum field
  theory. 3. Scalar quartic couplings,''
  Nucl.\ Phys.\  B {\bf 249} (1985) 70.

\bibitem{Abbott:1980hw}
  L.~F.~Abbott,
  ``The Background Field Method Beyond One Loop,''
  Nucl.\ Phys.\ B {\bf 185} (1981) 189.

\bibitem{Abbott:1981ke}
  L.~F.~Abbott,
  ``Introduction to the background field method,''
  Acta Phys.\ Polon.\  B {\bf 13} (1982) 33.

\bibitem{Mihaila:2012fm}
  L.~N.~Mihaila, J.~Salomon and M.~Steinhauser,
  ``Gauge coupling beta functions in the Standard Model to three loops,''
  Phys.\ Rev.\ Lett.\  {\bf 108} (2012) 151602
  [arXiv:1201.5868 [hep-ph]].

\bibitem{Mihaila:2012pz}
  L.~N.~Mihaila, J.~Salomon and M.~Steinhauser,
  ``Renormalization constants and beta functions for the gauge couplings
  of the Standard Model to three-loop order,''
  arXiv:1208.3357 [hep-ph].

\bibitem{Chetyrkin:2012rz}
  K.~G.~Chetyrkin and M.~F.~Zoller,
  ``Three-loop beta-functions for top-Yukawa and the Higgs
  self-interaction in the Standard Model,''
  JHEP {\bf 1206} (2012) 033
  [arXiv:1205.2892 [hep-ph]].

\bibitem{Hahn:2000kx}
  T.~Hahn,
  ``Generating Feynman diagrams and amplitudes with FeynArts 3,''
  Comput.\ Phys.\ Commun.\  {\bf 140} (2001) 418
  [arXiv:hep-ph/0012260].

\bibitem{Gorishnii:1989gt}
  S.~G.~Gorishnii, S.~A.~Larin, L.~R.~Surguladze and F.~V.~Tkachov,
  ``MINCER: program for multiloop calculations in quantum field theory
   for the schoonschip system,''
  Comput.\ Phys.\ Commun.\  {\bf 55} (1989) 381.

\bibitem{Tentyukov:1999is}
  M.~Tentyukov and J.~Fleischer,
  ``A Feynman diagram analyzer DIANA,''
  Comput.\ Phys.\ Commun.\  {\bf 132} (2000) 124
  [arXiv:hep-ph/9904258].

\bibitem{Actis:2006ra}
  S.~Actis, A.~Ferroglia, M.~Passera and G.~Passarino,
  ``Two-Loop Renormalization in the Standard Model. Part I: Prolegomena,''
  Nucl.\ Phys.\ B {\bf 777} (2007) 1
  [hep-ph/0612122].

\bibitem{Actis:2006rb}
  S.~Actis and G.~Passarino,
  ``Two-Loop Renormalization in the Standard Model Part II: Renormalization Procedures and Computational Techniques,''
  Nucl.\ Phys.\ B {\bf 777} (2007) 35
  [hep-ph/0612123].

\bibitem{Actis:2006rc}
  S.~Actis and G.~Passarino,
  ``Two-Loop Renormalization in the Standard Model Part III: Renormalization Equations and their Solutions,''
  Nucl.\ Phys.\ B {\bf 777} (2007) 100
  [hep-ph/0612124].


\bibitem{Christensen:2008py}
  N.~D.~Christensen and C.~Duhr,
  ``FeynRules - Feynman rules made easy,''
  Comput.\ Phys.\ Commun.\  {\bf 180} (2009) 1614
  [arXiv:0806.4194 [hep-ph]].

\bibitem{Semenov:2010qt}
  A.~Semenov,
  ``LanHEP - a package for automatic generation of Feynman rules from the
  Lagrangian. Updated version 3.1,''
  arXiv:1005.1909 [hep-ph].

\bibitem{Denner:1994xt}
  A.~Denner, G.~Weiglein and S.~Dittmaier,
  ``Application of the background field method to the electroweak standard
  model,''
  Nucl.\ Phys.\  B {\bf 440} (1995) 95
  [arXiv:hep-ph/9410338].

\bibitem{Tarasov:1980kx}
  O.~V.~Tarasov and A.~A.~Vladimirov,
  ``Three loop calculations in nonabelian gauge theories,''
JINR-E2-80-483.

\bibitem{'tHooft:1972fi}
  G.~'t Hooft and M.~J.~G.~Veltman,
  ``Regularization and Renormalization of Gauge Fields,''
  Nucl.\ Phys.\ B {\bf 44} (1972) 189.
  
\bibitem{Vermaseren:2000nd}
  J.~A.~M.~Vermaseren,
  ``New features of FORM,''
  arXiv:math-ph/0010025.

\bibitem{vanRitbergen:1998pn}
  T.~van Ritbergen, A.~N.~Schellekens and J.~A.~M.~Vermaseren,
  ``Group theory factors for Feynman diagrams,''
  Int.\ J.\ Mod.\ Phys.\  A {\bf 14} (1999) 41
  [arXiv:hep-ph/9802376].


\bibitem{Adler:1969gk}
  S.~L.~Adler,
  ``Axial vector vertex in spinor electrodynamics,''
  Phys.\ Rev.\  {\bf 177} (1969) 2426.


\bibitem{Bell:1969ts}
  J.~S.~Bell and R.~Jackiw,
  ``A PCAC puzzle: pi0 --> gamma gamma in the sigma model,''
  Nuovo Cim.\ A {\bf 60} (1969) 47.

\bibitem{Bardeen:1969md}
  W.~A.~Bardeen,
  ``Anomalous Ward identities in spinor field theories,''
  Phys.\ Rev.\  {\bf 184} (1969) 1848.

\bibitem{Bouchiat:1972iq}
  C.~Bouchiat, J.~Iliopoulos and P.~Meyer,
  ``An Anomaly Free Version of Weinberg's Model,''
  Phys.\ Lett.\ B {\bf 38} (1972) 519.

\bibitem{Gross:1972pv}
  D.~J.~Gross and R.~Jackiw,
  ``Effect of anomalies on quasirenormalizable theories,''
  Phys.\ Rev.\ D {\bf 6} (1972) 477.

\bibitem{Jegerlehner:2000dz}
  F.~Jegerlehner,
  ``Facts of life with gamma(5),''
  Eur.\ Phys.\ J.\ C {\bf 18} (2001) 673
  [hep-th/0005255].

\bibitem{Ferreira:1996ug}
  P.~M.~Ferreira, I.~Jack and D.~R.~T.~Jones,
  ``The three loop SSM beta functions,''
  Phys.\ Lett.\  B {\bf 387} (1996) 80
  [arXiv:hep-ph/9605440].

\bibitem{Jack:2004ch}
  I.~Jack, D.~R.~T.~Jones and A.~F.~Kord,
  ``Snowmass benchmark points and three-loop running,''
  Annals Phys.\  {\bf 316} (2005) 213
  [arXiv:hep-ph/0408128].

\bibitem{Harlander:2009mn}
  R.~V.~Harlander, L.~Mihaila and M.~Steinhauser,
  ``The SUSY-QCD beta function to three loops,''
  Eur.\ Phys.\ J.\ C {\bf 63} (2009) 383
  [arXiv:0905.4807 [hep-ph]].

\bibitem{Harlander:2007wh}
  R.~V.~Harlander, L.~Mihaila and M.~Steinhauser,
  ``Running of alpha(s) and m(b) in the MSSM,''
  Phys.\ Rev.\  D {\bf 76} (2007) 055002
  [arXiv:0706.2953 [hep-ph]].

\bibitem{Bauer:2008bj}
  A.~Bauer, L.~Mihaila and J.~Salomon,
  ``Matching coefficients for $\alpha_s$ and $m_b$ to $\mathcal{O}(\alpha_s^2)$ in the
  MSSM,''
  JHEP {\bf 0902} (2009) 037
  [arXiv:0810.5101 [hep-ph]].
\end{thebibliography}
\end{document}